\documentclass[prc,showpacs,showkeys,superscriptaddress,nofootinbib,floatfix,twocolumn]{revtex4}
\usepackage{color}
\usepackage{amsfonts}
\usepackage{amsbsy}
\usepackage{mathrsfs}
\usepackage{graphicx}
\def\lsim{\mathrel{\rlap{
\lower4pt\hbox{\hskip-3pt$\sim$}}
    \raise1pt\hbox{$<$}}}     
\def\gsim{\mathrel{\rlap{
\lower4pt\hbox{\hskip-3pt$\sim$}}
    \raise1pt\hbox{$>$}}}     
\def\scr#1{\mbox{\scriptsize #1}}
\def\1#1{{\bf #1}}
\usepackage{graphicx}
\begin{document}
\title{Dynamical Freeze-out in 3-Fluid Hydrodynamics}%
\author{V.N.~Russkikh}\thanks{e-mail: russ@ru.net}
\affiliation{Gesellschaft
 f\"ur Schwerionenforschung mbH, Planckstr. 1,
D-64291 Darmstadt, Germany}
\affiliation{Kurchatov Institute, Kurchatov
sq. 1, Moscow 123182, Russia}
\author{Yu.B.~Ivanov}\thanks{e-mail: Y.Ivanov@gsi.de}
\affiliation{Gesellschaft
 f\"ur Schwerionenforschung mbH, Planckstr. 1,
D-64291 Darmstadt, Germany}
\affiliation{Kurchatov Institute, Kurchatov
sq. 1, Moscow 123182, Russia}
\begin{abstract}
Freeze-out procedure accepted in the model of 3-fluid
dynamics (3FD) \cite{3FD} is analyzed. This procedure is formulated in
terms of drain terms in hydrodynamic equations. Dynamics of the
freeze-out is illustrated by 1-dimensional simulations.
It is demonstrated that the resulting freeze-out reveals a nontrivial
dynamics depending on initial conditions in the expanding
``fireball''. The freeze-out
front is not defined just ``geometrically'' on the condition of the
freeze-out criterion met but rather is a subject the fluid evolution.
It competes with the fluid flow and not always reaches the place where
the freeze-out criterion is met. Dynamics of the freeze-out
in 3D simulations is analyzed. It is demonstrated that the late stage
of central nuclear collisions at top SPS energies is of the form
of three (two baryon-rich and one baryon-free) fireballs 
separated from each other.
\end{abstract}
%
\maketitle

\section{Introduction}

Hydrodynamics is now a conventional approach to simulations
of heavy-ion collisions. Even
review papers \cite{Clare,Stoecker86,MRS91,Rischke98,Ruuskanen06} do not
comprise a complete list of numerous applications of this approach.
The hydrodynamics is applicable to description of hot and dense
stage of nuclear matter, when the mean free path is well shorter
than the size of the system. However, as expansion proceeds, the system
becomes dilute, the mean free path becomes comparable to
the system size, and hence the hydrodynamic calculation should be
stopped at some instant. All hydrodynamic calculations are
terminated by a freeze-out procedure, while these freeze-out
prescriptions are somewhat different in different models.

In the present paper we would like to describe in more detail the
freeze-out procedure accepted in recently developed 3FD model
\cite{3FD,3FDm}.
The 3FD model is designed for
simulating heavy-ion collisions in the energy range from 
BNL Alternating Gradient Synchrotron (AGS) to 
CERN Super Proton Synchrotron (SPS). 
Unlike the conventional hydrodynamics, where local instantaneous
stopping of projectile and target matter is assumed, a specific
feature of the dynamic 3-fluid description is a finite stopping
power resulting in a counter-streaming regime of leading
baryon-rich matter.  The basic idea of a 3-fluid approximation to
heavy-ion collisions \cite{3FD,3FDm} is that at each space-time
point a generally nonequilibrium
distribution function of baryon-rich
matter, can be represented as a sum of two
distinct locally equilibrated contributions,
initially associated with constituent nucleons of the projectile
(p) and target (t) nuclei. In addition, newly produced particles,
populating the mid-rapidity region, are associated with a ``fireball''
(f) fluid.  This model is a
straightforward extension of the 2-fluid model with radiation of
direct pions \cite{MRS91,MRS88,INNTS} and (2+1)-fluid model
\cite{Kat93,Brac97,Brac00a}. In particular, the 3FD model
allows a certain formation time for the fireball-fluid production, during
which the matter of the fluid propagates without interactions.

We have started our simulations \cite{3FD,3FDflow,3FDpt}
with a simple hadronic equation of state (EoS) \cite{gasEOS}.
This EoS is a natural reference point for any other more elaborate EoS.
The 3FD model turned out to be able to reasonably
reproduce a large body of experimental data \cite{3FD,3FDflow,3FDpt} in a wide
energy range from AGS to SPS. This was done with the unique set of
model parameters summarized in Ref. \cite{3FD}.
Problems were met in description of transverse flow \cite{3FDflow}.
The directed flow required a softer EoS at top AGS and SPS energies
(in particular, this desired softening may signal occurrence of a
transition into quark-gluon phase).

In particular, transverse-mass spectra of various hadrons were
reproduced \cite{3FD,3FDpt}. Experimental data on transverse-mass
spectra of kaons produced in central Au+Au \cite{E866} or Pb+Pb
\cite{NA49} collisions reveal peculiar dependence on the incident
energy. The inverse-slope parameter (so called effective
temperature) of these spectra at mid rapidity increases with
incident energy in the AGS energy domain and then saturates at the
SPS energies. In Refs. \cite{Gorenstein03,Mohanty03} it was assumed
that this saturation is associated with the deconfinement phase
transition. This assumption was indirectly confirmed by the fact
that microscopic transport models, based on hadronic degrees of
freedom, failed to reproduce the observed behavior of the kaon
inverse slope \cite{Bratkovskaya}. Hydrodynamic simulations of Ref.
\cite{Hama04} succeeded to describe this behavior provided the
incident-energy dependence of the freeze-out temperature has a very
similar shape to that of the corresponding kaon effective temperature.
Thus, the puzzle of kaon effective temperatures was just translated
into a puzzle of freeze-out temperatures.

In Ref. \cite{3FDpt} it was shown that dynamical description of freeze-out,
accepted in the 3FD model, naturally explains the  incident energy
behavior of inverse-slope parameters of transverse-mass spectra observed in
experiment. This freeze-out dynamics, effectively resulting in a
pattern similar to that of the dynamic liquid--gas transition, differs
from conventionally used  freeze-out schemes. This is the prime reason
why we would like to return to discussion of assumptions underlying this
prescription and present one-dimensional simulations, clarifying
consequences of this freeze-out.  It is
natural to start this discussion with a critical review of standard
assumptions of the freeze-out and recent developments in this field.

\section{Freeze-Out: Still Debated Problem}
\label{Freeze-Out}

The hydrodynamic simulation is terminated by a freeze-out
procedure. Though this method (as applied to high-energy physics)
was first proposed almost 50 years ago by Milekhin \cite{Milekhin},
this is a still debated problem. The method  was intuitively
clear and easily applicable. However, Cooper and
Frye \cite{Cooper} claimed that Milekhin's method violates the energy
conservation. To remedy the situation, they proposed their own
recipe, in which the observable spectrum of $a$-hadrons is calculated
as follows
\begin{eqnarray}
\label{Cooper-FO}
E \frac{d N_a}{d^3 p} =
\int_\Sigma d\sigma \
(p_\mu  n_\sigma^\mu) \ f_a (p,x)
\end{eqnarray}
where $\Sigma$ is a 3-dimensional hypersurface on which a certain criterion
of the freeze-out is met. Here integration runs over this hypersurface,
$n_\sigma^\mu$ is normal vector to the element $d\sigma$ of this
hypersurface, and $f_a (p,x)$ is an equilibrium distribution function
of $a$-hadrons
\begin{eqnarray}
\label{f-eq} f_a (p,x) =
\frac{g_a}{(2\pi)^3}%
\frac{1}{\exp\left\{\left(p_\mu u^\mu - \mu_a \right)/T\right\}\pm 1}
\end{eqnarray}
defined in terms of local thermodynamic and hydrodynamic quantities
on this freeze-out hypersurface: chemical potential $\mu_a(x)$,
temperature $T(x)$ and 4-velocity $u^\mu(x)$. Here $g_a$ is
degeneracy of the $a$ particle.

The Cooper--Frye recipe \cite{Cooper} is now extensively used in
hydrodynamic calculations, see, e.g.,
\cite{Ris95a,Kol99,HS95,Teaney01,Hirano07,Bass00,Per00,Non00,Hir02,Ham05,%
Nonaka06,Hirano06,Sat06}. However, it is not free of problems
neither. It gives negative contribution to the particle spectrum in
some kinematic regions in which the normal vector to the freeze-out
hypersurface is space-like, $p_\mu  n_\sigma^\mu < 0$. This
negative contribution corresponds to frozen out particles returning
to the hydro phase. Cut off of this negative contribution again
returns us to the violation of the energy conservation. To get rid
of this negative spectrum, there was proposed a modification of the
Cooper--Frye recipe based on a cut-J\"uttner distribution
\cite{Bugaev96,Neumann97,Csernai97,Bugaev99}. In this distribution
the part of the J\"uttner distribution that gave the negative
spectrum is simply cut off. To preserve the particle and energy
conservation, the rest of J\"uttner distribution is renormalized,
effectively resulting in a new temperature and chemical potential
(so called "freeze-out shock"). In fact, this cut-J\"uttner recipe
has no physical justification, except for practical utility.
Moreover, the cut-J\"uttner recipe is not supported by schematic
kinetic treatment \cite{Csernai99} of the transitional region from
hydro regime to that of dilute gas. 
Recently
there was proposed a new freeze-out recipe, a canceling-J\"uttner
distribution \cite{Csernai04}, which complies with results of
schematic kinetic treatment \cite{Csernai99}. It should be stressed
that this was precisely the schematic kinetic treatment. This
region, where the transition from highly collisional dynamics to the
collisionless one occurs, is highly difficult for the kinetic
treatment and hardly allows any justified simplifications.

All above considerations of the freeze-out process, including both
the original Cooper--Frye prescription and its improvements, proceed
from the following assumptions:
\begin{enumerate}
\def\theenumi{\Roman{enumi}}
    \item
"Decoupling" of matter from hydrodynamic regime happens on
{\it a continuous hypersurface} $\Sigma$.
    \item
This hypersurface is determined on the requirement that
{\it a certain criterion of the freeze-out is met}: e.g.,
temperature, energy density or baryon density reaches
a certain value.
    \item
After this "decoupling" particles {\it stream freely} to detectors.
\end{enumerate}
In fact, transition from highly collisional (hydro) regime to
collisionless one occurs in some finite 4-volume. Assumption (I)
is just an idealization---this 4-volume is shrunk to a
hypersurface. Conservation conditions
on such hypersurface are constructed in analogy with shock front in
hydrodynamics and result in the Cooper--Frye formula (\ref{Cooper-FO}).
However, the requirement that this surface is
continuous does not follow from anywhere. It is just an assumption.
For instance, if we assume a discontinuous hypersurface,
i.e. that consisting of tiny (infinitely small in continuum
limit) fragments with normal vectors
coinciding with local 4-velocity, $n_\sigma^\mu=u^\mu$, then
we return to the original Milekhin's method of the freeze-out.

The  Milekhin's method assumes that a hydro system freezes out by emitting tiny
fireballs of matter. Let $P^\mu_{\scr{tot}}$ is the total 4-momentum
of the system. Then at the first step of the freeze-out a tiny
droplet with the 4-momentum $\Delta P^\mu_{i}$ is emitted
\begin{eqnarray}
\label{Ptot+Pf}
 P^\mu_{\scr{tot}}=P^\mu_{\scr{fluid}}+\Delta P^\mu_{i},
\end{eqnarray}
where $P^\mu_{\scr{fluid}}$ is the 4-momentum of the still
hydro-evolving fluid. In terms of the energy-momentum tensor
$T^{\mu\nu}_{(i)}$, the
$\Delta P^\mu_{i}$ 4-momentum can be written out as
follows
\begin{eqnarray}
\label{Pf=int-T}
 \Delta P^\mu_{i} = \int_{\Delta V_i} dV \  T^{\mu 0}_{(i)} =
\int_{\Delta\Sigma_i} d\sigma \ T^\mu_{(i)\nu} n_\sigma^\nu,
\end{eqnarray}
where $\Delta V_i$ is the volume of the fireball in the reference
frame, where $T^{\mu 0}_{(i)}$ is considered. The last equality in
Eq. (\ref{Pf=int-T}) represents $\Delta P^\mu_{i}$ in the covariant
way, i.e. in terms a hypersurface element $\Delta\Sigma_i$ and the
normal vector to this element $n_\sigma^\mu$, cf.  Eq.
(\ref{Cooper-FO}). In particular,  Milekhin's choice consists in
$n_\sigma^\mu=u^\mu$. From representation (\ref{Pf=int-T}) it may
seem that relation between $\Delta P^\mu_{i}$ and $T^{\mu\nu}_{(i)}$
depends on $\Delta\Sigma_i$. This would imply that a proper
hypersurface element $\Delta\Sigma_i$ should be chosen to maintain
relation (\ref{Pf=int-T}). In fact, the r.h.s. of Eq.
(\ref{Pf=int-T}) is independent of $n_\sigma^\mu$. The formal proof
of that can be found, e.g., in Ref. \cite{Weinberg}. It is possible
to demonstrate it in a simpler way. Let us write down $T^{\mu
0}_{(i)}$ in terms of contributions of individual particles
\cite{Weinberg}
\begin{eqnarray}
\label{T=sum_part}
T^{\mu 0}_{(i)} ({\bf x},t) = \sum_n p^\mu_n (t) \ \delta^3
({\bf x}-{\bf x}_n (t)),
\end{eqnarray}
where $p^\mu_n (t)$ and ${\bf x}_n$ are the 4-momentum and the instant
position of the $n$th particle, respectively. Integrating expression
(\ref{T=sum_part}) over volume $\Delta V_i$, accordingly to
Eq. (\ref{Pf=int-T}), we arrive at
\begin{eqnarray}
\label{Pf=sun_n}
 P^\mu_{i} = \sum_n p^\mu_n (t).
\end{eqnarray}
Here spurious dependence on $\Delta\Sigma_i$ reveals itself in a
seeming dependence of the r.h.s. of Eq. (\ref{Pf=sun_n}) on the
synchronized time instant $t$, which really depends on the reference
frame and hence on $\Delta\Sigma_i$. Note that the $P^\mu_{i}$
quantity is assumed to be conserved, therefore the time dependence
of the r.h.s. of Eq. (\ref{Pf=sun_n}) is completely inappropriate.

Let us consider the r.h.s. of Eq. (\ref{Pf=sun_n}) in two reference
frames, i.e. on two hypersurface elements $\Delta\Sigma_i$ and
$\Delta\Sigma_i'$. The time synchronization depends on the reference
frame. Therefore, in the sums over particles
\begin{eqnarray}
\label{sun_n=sun_n'}
 \sum_n p^\mu_n (t) \quad\quad \mbox{and} \quad\quad \sum_n p'^\mu_n (t')
\end{eqnarray}
some $p^\mu_n (t)$ and $p'^\mu_n (t')$ may occur, which are not simply
related by the Lorentz transformation but are completely different
because the corresponding particles at the $t'$ instant have exercised
additional interactions (or vice versa, have not exercised all those
interactions) as compared to those completed to the $t$ instant.
And nevertheless two sums in Eq. (\ref{sun_n=sun_n'}) are equal, since
in each two-particle or multi-particle {\it point-like} interaction
the 4-momentum is conserved.
The point-like character\footnote{Action at distance in the
  relativistic case requires introduction of fields mediating this
  interaction. Then the field contribution should be also included in
$T^{\mu 0}$. For the sake of simplicity, we confine ourselves to the
point-like interaction.}
 of the interaction is of prime importance
here. If particles interact point-like, they change their momenta
simultaneously in any reference frame. Thus, the
r.h.s. of Eq. (\ref{Pf=sun_n}) is really independent of time $t$ and
hence of $\Delta\Sigma_i$.

\begin{figure}[thb]
\includegraphics[width=6.3cm]{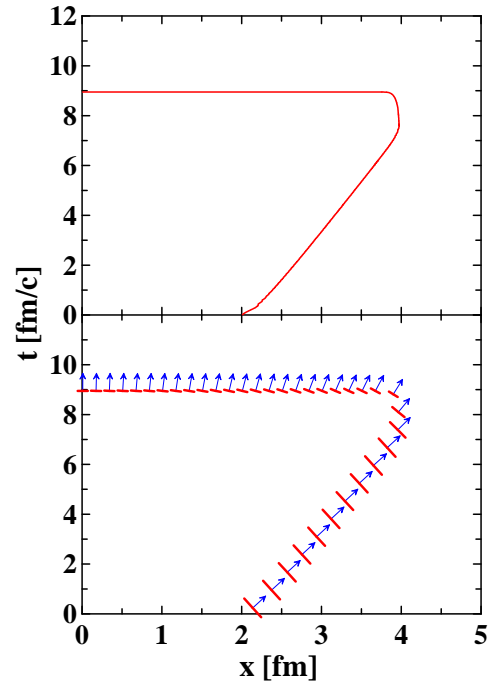}
\caption{(Color online)
Freeze-out hypersurface for hydrodynamic 
evolution the 1D
step-like slab of nuclear matter (see subsect. \ref{One-dimensional}). 
Initial conditions for this slab are constructed on the assumption
that they are formed by the shock-wave mechanism
in head-on collisions of two 1D slabs at
$E_{\scr{lab}}=$ 10 $A$ GeV. 
The upper panel displays the Cooper--Frye choice for the
hypersurface. The lower panel schematically illustrates Milekhin's
choice for the hypersurface. Arrows indicate local 4-velocities on
this hypersurface. 
}
\label{fig0}
\end{figure}

In view of Eqs. (\ref{Ptot+Pf}) and (\ref{Pf=int-T}),
upon completion of the freeze-out process, we have
\begin{eqnarray}
\label{Ptot=sumPf}
 P^\mu_{\scr{tot}}=\sum_{i} \Delta P^\mu_{i} =
\sum_{i} \int_{\Delta\Sigma_i} d\sigma \ T^\mu_{(i)\nu}
n_\sigma^\nu
=
\int_{\Sigma} d\sigma \ T^\mu_\nu n_\sigma^\nu,
\end{eqnarray}
where the hypersurface $\Sigma$ consists of elements $\Delta\Sigma_i$.
As we have seen, this 4-momentum conservation does not depend on the
choice of hypersurface elements $\Delta\Sigma_i$.  Milekhin's
choice is $n_\sigma^\mu=u^\mu$ and results in a discontinuous
hypersurface. The Cooper--Frye choice proceeds from requirement of
continuity of the $\Sigma$ hypersurface. Difference between these two
choices is illustrated in Fig. \ref{fig0}. The lower panel of
Fig. \ref{fig0} shows a schematic structure of Milekhin's
hypersurface. In practical calculations the fragments of Milekhin's
hypersurface are so tiny that the whole hypersurface looks like in
upper panel  of Fig. \ref{fig0}, however, with normal vector to each
tiny fragment coinciding with the 4-velocity. 

Therefore, Milekhin's method in fact conserves the energy, but to
see it one should consider it on a discontinuous hypersurface. The
baryon number conservation can be demonstrated in a similar way. The
fact that different single-particle distributions (i.e.
Cooper--Frye, cut-J\"uttner with renormalization,
canceling-J\"uttner, and even Milekhin's distributions) provide all
the required conservation laws and at the same time produce
different particle spectra, cf.  Eq. (\ref{Cooper-FO}) and Refs.
\cite{Cooper,Gorenstein84}, only increases ambiguity of freeze-out
consequences and indicate the barest necessity for further studies
of the freeze-out. Such studies of the freeze-out in a finite
4-volume (a finite space-like layer \cite{Csernai05}) are now in
progress.

From the practical point of view, assumption (II) means that
we should first run the hydro calculation without any freeze-out
and only after that look for a hypersurface, where the freeze-out
criterion is met. This hypersurface is determined as if the
hydrodynamic system is not affected by the freeze-out.
This procedure indeed results in a continuous
hypersurface, which could justify assumption (I). 
%
%
The method of ``continuous
emission'' \cite{Sinyukov02} offers a more consistent way of
performing freeze-out. This method considers a continuous
emission of particles from a finite volume, governed by their mean
free paths. In this approach
the freeze-out process looks like an evaporation
(or fragmentation, on account of final-size grid) first from the
system surface and then as a volume fragmentation of the system
residue.
The particle emission from surface layer of
the mean-free-path width may hint at a discreetness of the
hypersurface arising when this width is shrunk to zero.

In particular, the continuous-emission
 mechanism implies that dynamics of this
evaporation, characterized by its own rate, competes with the
hydrodynamic expansion. Therefore, freeze-out front may not reach
the places, where, e.g., the energy density reaches the critical value.
This is already a consequence of real freeze-out dynamics.
Unfortunately, this method is very difficult for the numerical
implementation because probability for a particle to leave the
hydrodynamic system depends not only on the past but
also on the future evolution of this
system, since particle emission occurs from time-evolving system.

Assumption (III) is also an approximation to reality.
This was the reason why cascading was applied after the
hydrodynamic freeze-out in Refs. \cite{HS95,Bass00,Hirano06,Nonaka06}.
This cascading allowed,
in particular, to reproduce a two-slope form of transverse-mass
spectra \cite{HS95} and a correct value of elliptic flow
\cite{Teaney01,Hirano07}. 
Inclusion of some inelastic channels in this cascading
may be important for proper reproduction
of multiplicities, e.g., the $K^-$ multiplicity \cite{3FD}.
%
%
However, even this cascading
is not enough. A mean-field cascading is really needed.
The reasons for this are as follows.
First of all,
the matter at the freeze-out instant is still dense enough, such that
a part of energy is accumulated in collective mean fields.
This mean-field energy should be released before calculating observables.
%
%
In the presently discussed 3FD model \cite{3FD} we do this 
at the freeze-out stage by recalculating 
thermodynamic quantities in terms the hadronic gas EoS rather than a
nontrivial EoS used in the hydro computation. 
%

\section{Freeze-out in 3FD}
\label{Freeze-out in 3FD}

The freeze-out scheme adopted in the 3FD model is an attempt to modify
and improve
the standard freeze-out procedure in certain aspects
rather than a final solution of the
freeze-out problem. Let us start with criteria of the freeze-out. We
formulate these criteria in terms of energy density, which is a universal
quantity applicable both at very high energies (instead of temperature)
and at low energies (instead of baryon density).
\\\\
{\bf (i)}
The freeze-out criterion we use is
\begin{eqnarray}
\label{FOcriterion1}
\varepsilon < \varepsilon_{\scr{frz}},
\end{eqnarray}
where
\begin{eqnarray}
\label{eps_tot}
\varepsilon = u_\mu T^{\mu\nu} u_\nu
\end{eqnarray}
is the total energy density of all three fluids in the proper reference
frame, where the composed matter is at rest. This total energy density is
defined in terms of the total energy--momentum tensor
\begin{eqnarray}
\label{T_tot}
T^{\mu\nu} \equiv
T^{\mu\nu}_{\scr p} + T^{\mu\nu}_{\scr t} + T^{\mu\nu}_{\scr f}
\end{eqnarray}
being the sum over energy--momentum tensors of separate fluids, and
the total collective 4-velocity of the matter
\begin{eqnarray}
\label{u-tot}
u^\mu = u_\nu T^{\mu\nu}/(u_\lambda T^{\lambda\nu} u_\nu).
\end{eqnarray}
Note that definition (\ref{u-tot}) is, in fact, an equation
determining $u^\mu$. In general, this $u^\mu$ does not coincide with 
4-velocities of separate fluids. 
This definition of the collective 4-velocity is in the spirit of the
Landau--Lifshitz approach to viscous relativistic hydrodynamics
\cite{Land-Lif}. 
Only the formed (to the time
instant of consideration) part of the f-fluid is taken into account in
$T^{\mu\nu}_{\scr f}$ of Eq. (\ref{T_tot}). To the end of
the freeze-out process all f-fluid turns out to be formed and hence
frozen out.

In the present simulations we use the value
$\varepsilon_{\scr{frz}}= 0.4$ GeV/fm$^3$ as the
critical freeze-out energy density, with the exception of low incident
energies $E_{\scr{lab}}$, for which we use lower values: 
$\varepsilon_{\scr{frz}}(2A \mbox{ GeV}) = 0.3$ GeV/fm$^3$ and 
$\varepsilon_{\scr{frz}}(1A \mbox{ GeV}) = 0.2$ GeV/fm$^3$. 
\\\\
{\bf (ii)}
In order to prevent freeze-out of initial cold nuclei, we apply the additional
criterion
\begin{eqnarray}
\label{FOcriterion2}
u_\mu \partial^\mu \varepsilon < 0
\quad \mbox{at the system surface},
\end{eqnarray}
i.e. at the boarder of matter with vacuum, which coincides with the
freeze-out front. In the frame, where the freeze-out front is at rest,
$\partial_t \varepsilon=0$, condition (\ref{FOcriterion2})
reduces to ${\bf u} {\bf \nabla} \varepsilon < 0 $,
which demands that collective velocity of the matter is directed outside
the system.
To meet this condition, we in fact start the freeze-out procedure only
at the expansion stage of the collision (see next subsect.). 
\\\\
{\bf (iii)}
A very important feature of our freeze-out procedure is an anti-bubble
prescription, preventing formation of bubbles of frozen-out matter
inside the dense matter still hydrodynamically evolving. 
The matter is allowed to be frozen out (provided two above
criteria are met) only if \\
{\bf (a)} either the matter is located near the boarder with
vacuum (this piece of matter gets locally frozen out) \\
{\bf (b)} or the maximal value of the total energy density in the system
is less than $\varepsilon_{\scr{frz}}$
\begin{eqnarray}
\label{FOcriterion3}
\max \varepsilon \leq \varepsilon_{\scr{frz}}
\end{eqnarray}
(the whole system gets instantly frozen out). \\

Criterion {\bf (iiib)} is convenient for numerical implementations while
does not look quite physical. From physical point of view, it would be
preferable to change $\max \varepsilon$ to the energy density
averaged over the system, $\langle \varepsilon\rangle$.
In view of the discussion below, such a substitution will not  change
the qualitative pattern of the freeze-out, however, can somewhat affect
quantitative results at AGS energies.

Before the instant of the global freeze-out, cf. {\bf (iiib)},
above freeze-out
criteria can be summarized in terms of dynamic equations
   \begin{eqnarray}
   \label{eq1}
   \partial_{\mu} J_{\alpha}^{\mu} =
   \Theta_{\scr{frz}} J_{\alpha}^{\mu} \partial_{\mu} \Theta_{\scr s},
   \end{eqnarray}
   \begin{eqnarray}
   \partial_{\mu} T^{\mu\nu}_{\alpha}  =
(\mbox{Friction})^\nu  +
\Theta_{\scr {frz}} T^{\mu\nu}_{\alpha} \partial_{\mu} \Theta_{\scr s},
   \label{eq2}
   \end{eqnarray}
where $J_{\alpha}^{\mu}$ is the baryon current of the $\alpha$ fluid,
$\alpha =$ p, t or f (i.e. projectile, target or fireball), note that
$J_{\scr f}^{\mu}\equiv 0$.
Here $(\mbox{Friction})^\nu$ stands for interaction terms between fluids,
the explicit form of which is not important here. $\Theta_{\scr s}$ is a step
function at the system surface, which takes into account criterion
{\bf (iiia)}.
1-dimensional simulations of the freeze-out {\bf (i)-(iiia)} show that
it results
in discontinuity of $\varepsilon$ (and other quantities) at
the system surface. This discontinuity is numerically smeared out only
to the extent of the finite grid step.
Therefore, in analytic equations
the step function $\Theta_{\scr s}$ can be represented by a sharp
step function
   \begin{eqnarray}
   \label{theta-s}
   \Theta_{\scr s} = \Theta (\varepsilon-\delta)
   \end{eqnarray}
with $\delta\to +0$. The function
   \begin{eqnarray}
   \label{theta-frz}
   \Theta_{\scr {frz}} =
   \Theta (\varepsilon_{\scr{frz}}-\varepsilon^s)
   \Theta (-u_\mu \partial^\mu \varepsilon)
   \end{eqnarray}
with $\varepsilon^s$ being the value of $\varepsilon$
at matter side of the surface discontinuity, takes into account conditions
{\bf (i)} and {\bf (ii)} of the freeze-out.

Let freeze-out conditions {\bf (i)} and {\bf (ii)} be met, i.e.
$\Theta_{\scr {frz}}=1$.
To verify that Eqs. (\ref{eq1})--(\ref{theta-frz}) actually correspond to
the above described scheme {\bf (i)-(iiia)}, let us consider these equations
for the stationary situation
in the reference frame, where the freeze-out front is at rest, i.e.
$\partial_{t} \Theta_{\scr s} = 0$. For the sake of convenience, let us
associate the freeze-out front with the $x=0$ plane. Then the only nonzero
component of the $\partial_{\mu} \Theta_{\scr s}$ 4-vector is
$\partial_{x} \Theta_{\scr s} = -\delta (x)$ (the hydrodynamic matter occupies
the $x<0$ semi-space), and  Eqs. (\ref{eq1})--(\ref{eq2}) take the form
   \begin{eqnarray*}
   \partial_{x} J_{\alpha}^{x} &=&
    -J_{\alpha(s)}^{x} \delta (x),
\\
%
   \partial_{x} T^{x\nu}_{\alpha}  &=&
- T^{x\nu}_{\alpha(s)} \delta (x),
   \end{eqnarray*}
where symbol $(s)$ in the subscript indicate that the value is taken
at the matter side of the freeze-out front. The latter is the effect
of $\delta\to +0$ introduced in Eq. (\ref{theta-frz}). We have also
omitted friction forces in the r.h.s. of Eq. (\ref{eq2}), since they are
really unimportant at the freeze-out stage. Integrating above equations
over small $\Delta x$ around $x=0$, we arrive at
   \begin{eqnarray*}
   \Delta J_{\alpha}^{x}&\equiv&
J_{\alpha(x>0)}^{x}-J_{\alpha(s)}^{x} =  -J_{\alpha(s)}^{x},
\\
   \Delta T^{x\nu}_{\alpha}&\equiv&
T^{x\nu}_{\alpha(x>0)}- T^{x\nu}_{\alpha(s)} = - T^{x\nu}_{\alpha(s)}.
   \end{eqnarray*}
Thus,
terms $\propto \partial_{\mu} \Theta_{\scr s}$ in the r.h.s. of Eqs.
(\ref{eq1}) and (\ref{eq2}) play role of sinks, which remove matter from
hydrodynamic  evolution, making the hydrodynamic quantities
$J_{\alpha(x>0)}$ and $T_{\alpha(x>0)}$  to be zero after the freeze-out front.

This kind of freeze-out is similar to the model of ``continuous
emission'' proposed in Ref. \cite{Sinyukov02}. There the particle emission
occurs from a surface layer of the mean-free-path width. In our case the
physical pattern is  similar, only the mean free path is shrunk to zero.


The above discussion concerns only the first part of the freeze-out procedure,
i.e. the application of freeze-out conditions. The second part consists in
calculation of spectra of observable particles:
\\\\
{\bf (iv)} Milekhin's method  \cite{Milekhin},
defined on a discontinuous hypersurface was used for calculation of
observables, cf. sect. \ref{Freeze-Out}. \\

Thus, this freeze-out is similar to evaporation of particles from a free 
surface of the system followed by explosion of the fluid residue, if criterion 
(\ref{FOcriterion1}) is met in the whole finite volume of this residue.

\subsection{Physical pattern of the freeze-out}
\label{Physical}

The physical pattern behind this freeze-out resembles the process of 
expansion of compressed and heated classical fluid into vacuum. Physics of this 
process is studied both experimentally and theoretically
\cite{evap76,evap82,evap92,evap99}. Evaporation from free surface of normal 
(not superheated) fluid is a very slow process. Accordingly, the freeze-out of 
matter of high density ($\varepsilon > \varepsilon_{\scr{frz}}$) is suppressed in
our model. 
During expansion the fluid becomes more rarefied, still remaining quite hot. Thus, 
the fluid becomes superheated at $\varepsilon < \varepsilon_{\scr{frz}}$. 
It occurs first at the periphery of the system, 
which is first affected by the decompression wave. Evaporation from free surface 
of superheated fluid is already a fast process. Accordingly, the freeze-out 
is allowed at $\varepsilon < \varepsilon_{\scr{frz}}$.

Situations are possible, when freeze-out criterion (\ref{FOcriterion1})
is met in the whole slab near the free surface rather than only at the surface. 
Such situations are illustrated in Subsects. \ref{One-dimensional} and 
\ref{3D simulations}. Here we have a choice either to instantaneously 
freeze out this whole near-surface slab or to wait until the freeze-out front 
will gradually traverse this slab (if ever). Making this choice, we rely on 
results of experiments on evaporation from superheated fluids. It was shown (see, 
e.g., Ref. \cite{evap99}) that the evaporation front propagates with respect 
to fluid not faster than with the speed of sound. Precisely this choice is 
realized in our model by means of condition {\bf (iiia)} or alternatively by 
dynamic equations (\ref{eq1}) and (\ref{eq2}). Thus, the freeze-out front may 
stay at essentially lower energy densities than $\varepsilon_{\scr{frz}}$ 
because supersonic fluid expansion prevents it from reaching the region, where 
$\varepsilon = \varepsilon_{\scr{frz}}$.

Physically it implies that a particle is evaporated ("frozen-out") only if it 
escapes from the system without collisions. Thus, its mean free path 
($\lambda_{\scr{mfp}}$) should be 
larger than its path to the free surface (with due account of the future 
evolution of the fluid). Precisely this criterion is applied in the 
model of ``continuous emission'' \cite{Sinyukov02}. In our simplified version 
of the ``continuous emission'', $\lambda_{\scr{mfp}}=0$ in the fluid phase and 
$\lambda_{\scr{mfp}}\to\infty$ in the gas phase. Therefore, a particle can escape 
only from the free surface, which cannot move inward the system faster than  
with the speed of sound \cite{evap99}.

The only exception from this rule we do at the final stage of the freeze-out. 
As it was observed in experiments with classical fluids  (see, 
e.g., Ref. \cite{evap82}), a fluid transforms into 
gas by explosion, if it is strongly superheated all over its volume. Therefore, 
at the final stage of the freeze-out, when criterion (\ref{FOcriterion1}) is met 
in the whole volume of the fluid residue, we assume that the whole residue becomes 
frozen out simultaneously (condition {\bf (iiib)}).

Of course, criterion (\ref{FOcriterion1}) is not universal. In particular, it is 
not applicable to the cold nuclear matter, which has $\varepsilon\approx$ 0.15 
GeV/fm$^3$ in its ground state. Additional condition (\ref{FOcriterion2}), 
preserving the cold nuclear matted from being frozen out, is directly 
connected with this fact. We hope that criterion (\ref{FOcriterion1}) is 
good enough for a restricted domain of the phase diagram, where freeze-out 
of hot nuclear matter really occurs.

In order to further clarify our freeze-out scheme it is useful to
consider the way it was implemented in the numerical scheme.

\subsection{Numerical implementation of the freeze-out}
\label{Numerical}

The numeric scheme of the 3FD code is based
on the modified particle-in-cell method \cite{Roshal81,FCT-PIC},
which is an extension of the scheme first applied in Los-Alamos
\cite{Harlow76}. 
In the particle-in-cell method each time step of the computation
consists of three stages: (I) Eulerian stage, (II) Lagrangian stage,
and (III) 
transformation from the frame of computation to
the local rest frame of fluids in order to
calculate thermodynamic quantities and flow velocities (as
applied to the freeze-out, it is described in point {\bf (c)} 
of this subsect.). 


The transfer of energy--momentum due to pressure gradients,
friction between fluids and production of the fireball fluid is
computed on the spatially fixed grid (so called Eulerian stage of the
scheme). The convective transfer of the baryonic charge, energy
and momentum is performed at the Lagrangian stage of the scheme. 
At this stage the matter is represented by an
ensemble of  Lagrangian particles which accumulate all the
energy, momentum and baryon charge of the system. 
At the time step (let it be ``1''), when the freeze-out has not
started yet,  
the total energy--momentum ($P^\mu_{\scr{tot}}$) and baryon charge
($B_{\scr{tot}}$) are presented by sums over these particles
\begin{eqnarray}
\label{P-test-part} 
\hspace*{-4mm} 
P^\mu_{\scr{tot}} &=& \sum_{i\alpha} \Delta P^\mu_{i\alpha} (t_1)
=  \sum_{i\alpha} {\Delta V_{i\alpha}(t_1)} \  
T^{\mu 0}_{(i\alpha)}(t_1), 
\\
\label{B-test-part} 
\hspace*{-4mm} 
B_{\scr{tot}} &=& \sum_{i\alpha} \Delta B_{i\alpha}(t_1)
=  \sum_{i\alpha} {\Delta V_{i\alpha}(t_1)} \  
J^{0}_{(i\alpha)}(t_1), 
\end{eqnarray}
where $\Delta P^\mu_{i\alpha}(t_1)$ and $\Delta B_{i\alpha}(t_1)$ are
respectively the energy--momentum and baryon charge of an $i$th
particle belonging to the $\alpha$-fluid. These sums run over both
formed and still unformed particles of the f-fluid. 
In fact, baryon charges of
all particles belonging to a baryon-rich fluid are taken to be constant
and equal, 
$\Delta B_{ip}(t)=b_p$ and $\Delta B_{it}(t)=b_t$ ($b_p=b_t$ if nuclei
are identical), while for the f-fluid 
$\Delta B_{if}(t)=0$. Each Lagrangian particle is also characterized
by a volume $\Delta V_{i\alpha}(t_1)$. Therefore, the above quantities of
the particle are expressed in terms of respective densities  
$T^{\mu 0}_{(i\alpha)}$ and $J^{0}_{(i\alpha)}$, as 
it is indicated in Eqs. (\ref{P-test-part}) and (\ref{B-test-part}). 
Simulation is performed in the frame of equal velocities of
colliding nuclei. Hence, all the quantities in
Eqs. (\ref{P-test-part}) and (\ref{B-test-part}) are related to this
frame. 
Eqs. (\ref{P-test-part}) and (\ref{B-test-part}) imply that all the matter
participates in dynamical evolution before the freeze-out starts: 
\begin{eqnarray}
\label{PB-part-start} 
P^\mu_{\scr{dyn}} (t_1) = P^\mu_{\scr{tot}},   
\quad\quad
B_{\scr{dyn}} (t_1) = B_{\scr{tot}}.   
\end{eqnarray}
Here $P^\mu_{\scr{dyn}}$ and $B_{\scr{dyn}}$ are
respectively total energy--momentum and baryon charge participating in
dynamical evolution. 
We avoid term ``hydrodynamic evolution'' because a part of the
f-fluid may be still unformed.

In the present scheme the Lagrangian particle has a profile function
of the form and size of the grid cell with uniform distribution of
densities. Therefore, in 3D simulations a single Lagrangian 
particle contributes to 8 cells on the grid, with which it overlaps. 
These spatially extended particles make the scheme smoother and hence
more stable. In Refs. \cite{3FD,MRS91} this numerical scheme is described
in more detail. 
\\

\noindent {\bf (a)} 
To roughly meet condition {\bf (ii)}, the freeze-out
procedure is started only
at the expansion stage of the collision, i.e. after the time interval
required for nuclei to traverse each other, provided they keep their initial
velocities. In the c.m. frame of two identical nuclei this time delay is
$\Delta t_{\scr{frz}}=D_A/(\gamma_{\scr{cm}} v_{\scr{cm}})$, where 
$D_A/\gamma_{\scr{cm}}$ is the Lorentz contracted diameter of the nucleus, 
$\gamma_{\scr{cm}}$ and $v_{\scr{cm}}$ are the $\gamma$ factor and the
initial velocity of the nucleus in the c.m. frame, respectively.
\\

\noindent {\bf (b)} 
The freeze-out criterion (\ref{FOcriterion1}) is checked at the
Lagrangian stage of each time step (let it be $n$th step). 
In the f-fluid only those particles are considered which have been
formed to the time instant $t_n$. For each Lagrangian particle 
it is checked in all cells, which
overlap with this considered particle, i.e. in 8 cells. 
%

If the freeze-out criterion is met in {\em all
these 8 cells} and if at least one of these cells is ``empty''
(i.e. contains no centers of any Lagrangian particles), then this
considered Lagrangian  particle 
is counted as frozen out. This is realization of condition {\bf   (iiia)}. 
This frozen-out Lagrangian particle is removed from
further hydrodynamic evolution. By doing this we remove respective
portions from sums (\ref{P-test-part}) and (\ref{B-test-part}): 
\begin{eqnarray}
\label{P-test-part-rem} 
\hspace*{-4mm} 
P^\mu_{\scr{dyn}} (t_n) &=& P^\mu_{\scr{dyn}} (t_{n-1})  - 
\hspace*{-2mm} 
\sum_{i\alpha\scr{ frozen out at }t_n} 
\hspace*{-6mm} 
\Delta P^\mu_{i\alpha} (t_n),
\\
\label{B-test-part-rem} 
\hspace*{-4mm} 
B_{\scr{dyn}} (t_n) &=& B_{\scr{dyn}} (t_{n-1})   - 
\hspace*{-2mm} 
\sum_{i\alpha\scr{ frozen out at }t_n} 
\hspace*{-6mm} 
\Delta B_{i\alpha}(t_n).  
\end{eqnarray}
Only parts $P^\mu_{\scr{dyn}}(t_n)$ and $B_{\scr{dyn}}(t_n)$ 
are kept in further dynamic evolution, unlike the conventional
Cooper--Frye method.  
It is important to mention that had we kept this frozen-out particle
in the dynamic evolution, its energy--momentum content would be
changed at later time steps, and hence a part of its energy--momentum
would be gained by other particles. Then we would face problems with the 
energy--momentum conservation, if we wanted to use the $\Delta
P^\mu_{i\alpha}$ quantity at the instant of its freeze-out for
calculation of observable spectra.

If the freeze-out criterion is met in {\em all
cells of the system}, all Lagrangian particles are counted as frozen out at
this time step, as it is required by condition {\bf   (iiib)}. 
This is the end of the freeze-out process. 
\\

\noindent {\bf (c)} 
{\it When the freeze-out process is over, i.e. $P^\mu_{\scr{dyn}}=0$
  and $B_{\scr{dyn}}=0$, we are left with  
ensemble of  frozen-out Lagrangian particles which precisely
obeys conservation laws:} 
\begin{eqnarray}
\label{P-test-part-frz} 
P^\mu_{\scr{tot}} &=& \sum_{i\alpha} \Delta P^\mu_{i\alpha} 
(t_{i\alpha\scr{ frozen-out}}), 
\\
\label{B-test-part-frz} 
B_{\scr{tot}} &=& \sum_{i\alpha} \Delta B_{i\alpha}
(t_{i\alpha\scr{ frozen-out}}). 
\end{eqnarray}
Here the summation runs over frozen-out (at different time instants 
$t_{i\alpha\scr{ frozen-out}}$) particles, unlike sums
(\ref{P-test-part}) and (\ref{B-test-part}), where this summation is
associated with a fixed time instant. 
These frozen-out particles are precisely  those droplets mentioned in
Eq. (\ref{Ptot+Pf}).

A frozen-out $i\alpha$-particle is still characterized by five 
hydrodynamic quantities, $J^0_{i\alpha}$ and  $T^{\mu 0}_{i\alpha}$,
and volume $\Delta V_{i\alpha}$, all these in the reference frame of
computation. For the calculation of spectra we need distribution
function formulated in terms of thermodynamic quantities: temperature
($T^{i\alpha}$), baryon ($\mu_b^{i\alpha}$)
and strange ($\mu_s^{i\alpha}$) chemical potentials, and 
hydrodynamic 4-velocity ($u^\mu_{i\alpha}$), cf. Eqs.
(\ref{Cooper-FO}) and (\ref{f-eq}). In fact, this recalculation of the 
hydrodynamic quantities into thermodynamic ones is performed at each
time step of the scheme based on the nongas EoS (involving some
mean fields) accepted in the calculation. 
This EoS 
is not suitable for calculation of the spectrum of observable
particles. First we should release the energy stored in mean
fields. To do this, we calculate $T^{i\alpha\scr{(gas)}}$, 
$\mu_b^{i\alpha\scr{(gas)}}$, $\mu_s^{i\alpha\scr{(gas)}}$ 
and $u^\mu_{i\alpha\scr{(gas)}}$ 
based on {\it the hadronic gas EoS} and proceeding from
conservations of total energy--momentum, baryon and strange charges
in the frozen-out particle. 
\\

\noindent {\bf (d)} 
At this stage we are still free to choose either
Cooper--Frye or Milekhin's scheme to calculate observables, as it was
argued in Eqs. (\ref{Ptot+Pf})--(\ref{Ptot=sumPf}). 
We use  Milekhin's method  \cite{Milekhin},
defined on a discontinuous hypersurface consisting of tiny fragments
coinciding with volumes of frozen-out Lagrangian particles $\Delta
V_{i\alpha}$.  
The observable spectrum of hadrons is calculated as follows
\begin{eqnarray}
\label{FOspectrum} E \frac{d N}{d^3 p} = \sum_{i\alpha}
V_{i\alpha}^{\scr{(proper)}}
p_\mu u^\mu_{i\alpha\scr{(gas)}}
f_{i\alpha\scr{(gas)}} (p)
\end{eqnarray}
where $V_{i\alpha}^{\scr{(proper)}}$ is the volume of the
$i\alpha$-particle in its rest frame, the sum runs over all
frozen-out particles of all fluids,
$f_{i\alpha\scr{(gas)}} (p)$ is the equilibrium distribution function
defined already in terms of local  {\it gas} thermodynamic 
($T^{i\alpha\scr{(gas)}}$, $\mu_b^{i\alpha\scr{(gas)}}$ and 
$\mu_s^{i\alpha\scr{(gas)}}$) and
hydrodynamic ($u^\mu_{i\alpha\scr{(gas)}}$) quantities,  
cf. Eqs. (\ref{Cooper-FO}) and (\ref{f-eq}).
\\

In this prescription the baryon number and energy-momentum are
precisely conserved by construction.  
It is worthwhile to
mention that both the Cooper--Frye and Milekhin methods possess the
same main problem: they both do not reject contributions of
frozen-out hadrons returning into the hydrodynamic phase 
and do this to precisely the same extent, see Appendix \ref{Problem}. In
particular, in our calculation this problem probably reveals itself in
the failure of reproduction of the pion directed flow
\cite{3FD,3FDflow}. The advantage of Milekhin's method is just
practical: with the exception of the pion directed flow it quite
successfully  works in the 3FD model. The canceling-J\"uttner recipe
\cite{Csernai04} overcomes the returning-hadrons problem of the
above methods. It would be of interest to apply it in the 3FD model.

\subsection{One-dimensional simulations}
\label{One-dimensional}

\begin{figure}[thb]
\includegraphics[width=7.9cm]{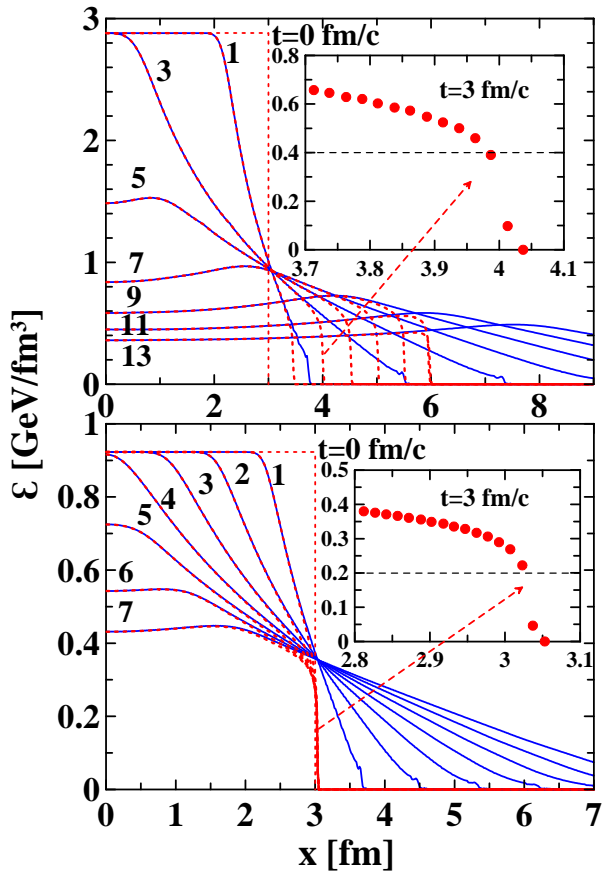}
\caption{(Color online)
Evolution of the energy density during the decay of the 1D
step-like slab of nuclear matter. Solid lines display calculations
without freeze-out, while dashed lines, with freeze-out.
Initial conditions for these slabs are constructed on the assumption
that they are formed by the shock-wave mechanism
in head-on collisions of two 1D slabs at
$E_{\scr{lab}}=$ 10 $A$ GeV for upper panel and $E_{\scr{lab}}=$ 2 $A$
GeV for lower panel. Subpanels show zoomed regions of the freeze-out
front at the fixed time instant $t=$ 3 fm/c in terms of the energy
density in separate cells (displayed by dots).
}
\label{fig1}
\end{figure}

In order to clarify physics described by
Eqs. (\ref{eq1})--(\ref{theta-frz}) let us consider 1D simulations
based on them.
In Fig. \ref{fig1} decays of  step-like slabs of nuclear matter are presented.
These simulations have been performed till the time instant of the
global  freeze-out, i.e. when criterion {\bf (iiib)} is met.
The same EoS
as that used in 3D simulations \cite{3FD,3FDflow,3FDpt}
is accepted in the present calculations.
First of all, we see that the freeze-out front is really step-like.
It is smeared only over two cells (independently of their size) due
to numeric scheme. Note also that
for this step-like initial geometry the supersonic flow of matter
($v_x > c_s$, where $c_s$ is the speed of sound)
always occurs beyond the initial boundary of the slab, while the flow
within the initial boundary is always subsonic ($v_x < c_s$).
This is important for
understanding results displayed in Fig. \ref{fig1}.

There are three important velocities in this problem: the hydrodynamic
velocity of the matter ($v_x$) at the position, where the freeze-out front
occurs, the speed of sound $c_s$, and the velocity $v_\varepsilon$ of
transfer of
the constant value of $\varepsilon = 0.4$ GeV/fm$^3$:
\begin{eqnarray}
\label{v-epsilon}
\varepsilon (v_\varepsilon t + \mbox{const},t) = 0.4 \ \mbox{GeV/fm}^3.
\end{eqnarray}
In fact, Eq. (\ref{v-epsilon}) is the equation for the hydrodynamic
characteristic curve related to the 0.4 GeV/fm$^3$ value of the energy
density. The freeze-out front, as defined by Eq. (\ref{eq1}) and
(\ref{eq2}), cannot propagate in the fluid medium
faster than with the local speed of sound, like any perturbation in the
hydrodynamics.

\begin{figure}[thb]
\includegraphics[width=7.9cm]{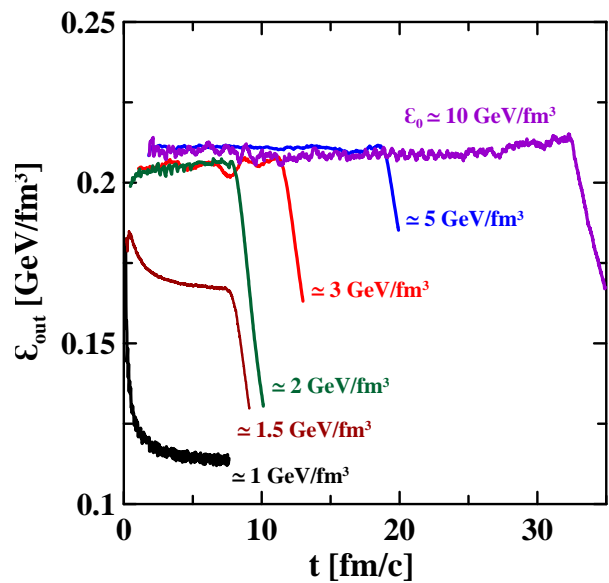}
\caption{(Color online)
Instant values of actual energy density of the frozen-out matter for
decays of nuclear-matter slabs with various initial 
conditions (labeled by initial energy densities $\varepsilon_0$).
Initial conditions  are constructed on the assumption
of the shock-wave mechanism
in head-on collisions of two slabs at
$E_{\scr{lab}}=$ 2, 4, 6, 10, 20 and 40 $A$ GeV (from bottom left to
top right).
The time evolution is displayed till the time instant of the global
freeze-out {\bf (iiib)}.
}
\label{fig1a}
\end{figure}

Different dynamic patterns of the freeze-out fronts displayed
in two panels of  Fig. \ref{fig1} are associated with different
relations between above three velocities. In the upper case of
$\varepsilon_0 \simeq 3$ GeV/fm$^3$  we have $v_\varepsilon>0$.
Hence, the point, where the freeze-out criterion (\ref{theta-frz})
starts to be met, is transferred farther and farther from the initial
system boundary, hence the system expands. In this region the flow is
supersonic: $v_x > c_s$. At the same time, the matter velocity with
respect to the characteristic velocity is\footnote{Note that
  velocities are added relativistically.}
$(v_x - v_\varepsilon)/(1- v_x v_\varepsilon) \leq c_s$, i.e. it is
less than the speed of sound. Therefore, 
the fluid flow does not carry the freeze-out front
away from the point of $\varepsilon = 0.4$ GeV/fm$^3$, where  the
freeze-out criterion (\ref{theta-frz}) is met. 
The freeze-out front
stays at this position, see the zoomed subpanel in the top panel of
Fig. \ref{fig1}. Since the matter within this freeze-out front 
is distributed over the range of $0<\varepsilon< 0.4$ GeV/fm$^3$, 
an average value for the actual energy
density of the frozen-out matter (let us denote it as
$\varepsilon_{\scr{out}}$) is approximately half of the freeze-out
jump,  i.e. $\varepsilon_{\scr{out}}\approx
\varepsilon_{\scr{frz}}/2 = 0.2$ GeV/fm$^3$ in this case, see  Fig.
\ref{fig1a}.

In the bottom panel of  Fig. \ref{fig1}, i.e.
$\varepsilon_0 \simeq 1$ GeV/fm$^3$,  we have
$v_\varepsilon<0$. Here, the point, where freeze-out criterion
(\ref{theta-frz}) starts to be met, cannot be reached by the freeze-out
front. Indeed, the freeze-out front is first formed beyond the initial
boundary of the slab, i.e. in the region of the supersonic flow of the
matter. Since it cannot be faster than the sound,
it cannot overcome the ``supersonic barrier'' in the back
direction in order to reach the $\varepsilon = 0.4$ GeV/fm$^3$ point.
It stays at the position of the ``supersonic barrier'' and does not
move. Therefore, the decaying and freezing-out system does not expand.
Moreover, the
front stays at the position of low energy density, $\varepsilon \simeq
0.2$ GeV/fm$^3$, see the zoomed subpanel in the bottom panel of
Fig. \ref{fig1}.
Therefore, the actual energy density of the frozen-out matter turns
out to be lower than $\varepsilon_{\scr{frz}}/2$ (as
in above case), i.e.
$\varepsilon_{\scr{out}}\approx (0.2$ GeV/fm$^3)/2 = 0.1$ GeV/fm$^3$,
see  Fig. \ref{fig1a}.

In Fig. \ref{fig1a} we see that $\varepsilon_{\scr{out}}$ remains
practically constant during the major period of the freeze-out. The
steep fall of $\varepsilon_{\scr{out}}$ just before the global
freeze-out occurs because the velocity $v_\varepsilon$ on the
characteristic curve (\ref{v-epsilon}) changes its sign. Then the
point, where $\varepsilon  = 0.4$ GeV/fm$^3$ is achieved, starts
rapidly move inwards the system, and hence the freeze-out front
remains at lower energy density.

Thus, we see that the freeze-out is not inseparably
linked with the  freeze-out energy density $\varepsilon_{\scr{frz}}$
but can occur at lower energy densities due to dynamical reasons.
At low
initial energy densities the freeze-out front stays at lower
energy densities than $\varepsilon_{\scr{frz}}$ and hence
the actual energy density of the frozen-out matter
$\varepsilon_{\scr{out}}$ turns out to be lower
than $\varepsilon_{\scr{frz}}/2$, see Fig. \ref{fig1a}. With the
initial energy density rising, the $\varepsilon_{\scr{out}}$ quantity
grows and gradually
reaches the $\varepsilon_{\scr{frz}}/2$ value and than approximately
saturates.

\begin{figure}[thb]
\includegraphics[width=7.9cm]{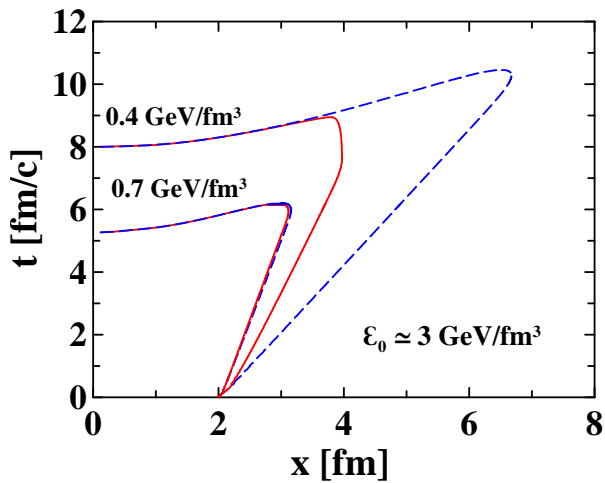}
\caption{(Color online)
Characteristic curves for hydrodynamic 
evolution the 1D step-like slab with 
initial conditions corresponding to the upper panel of Fig. \ref{fig1}.
Solid lines display characteristic curves, corresponding to 
$\varepsilon= 0.4$ and 0.7 GeV/fm$^3$, 
calculated with local freeze-out {\bf (iiia)} but without global
freeze-out {\bf (iiib)}.  
Dashed lines correspond to the calculation without freeze-out.
}
\label{fig1d}
\end{figure}

A standard procedure of performing freeze-out, which is applied in the
major part of hydrodynamic calculations now, proceeds in the following
way. The hydro calculation runs absolutely unrestricted. The freeze-out
hypersurface is determined by analyzing the resulting 4-dimensional
field of hydrodynamic quantities on the condition of the freeze-out
criterion met. In our case this would be the characteristic curve 
of $\varepsilon= 0.4$ GeV/fm$^3$ calculated  without freeze-out,
displayed in Fig. \ref{fig1d}.

In the 3FD model, the freeze-out criterion is checked continuously
during the simulation. If some parts of the hydro system meet all
criteria ({\bf (i)}, {\bf (ii)}, and either {\bf (iiia)} or {\bf
  (iiib)}), they decouple from the hydro calculation. 
The frozen-out matter escapes from the system, removing all the energy
and momentum accumulated in it, cf. Eq. (\ref{Ptot+Pf}). 
Therefore, it produces no
recoil to the rest of still hydrodynamic system. 
The removal of the matter affects the system evolution. 
This influence is illustrated in Fig. \ref{fig1d}. The 
$\varepsilon= 0.4$ GeV/fm$^3$ characteristic curves 
calculated with and without freeze-out turn out to be different. At
the same time  the $\varepsilon= 0.7$  GeV/fm$^3$ characteristic
curves, which lie quite deep inside the system, remain unaffected by the
freeze-out. The freeze-out hypersurface (i.e. curve in 1+1 dimensions)
for this case is presented in Fig. \ref{fig0}. It differs from the
corresponding characteristic curve because of the global freeze-out
which occurs at time instant $t\simeq$ 9 fm/c.

\begin{figure}[thb]
\includegraphics[width=7.2cm]{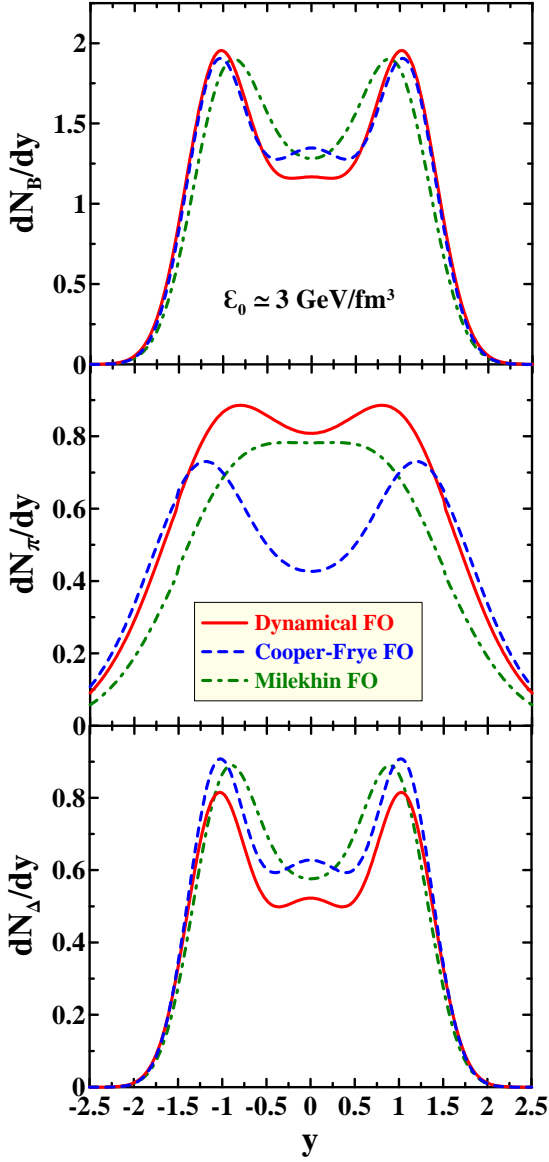}
\caption{(Color online)
Rapidity distributions of baryons (upper panel), thermal pions (middle panel) and 
$\Delta$-isobars (lower panel) calculated within three models of freeze-out: 
the model of dynamical freeze-out [Dynamical FO], 
[see Eqs. (\ref{FOcriterion1})--(\ref{FOspectrum})], 
Cooper--Frye freeze-out on characteristic curve 
(\ref{v-epsilon})
[Cooper--Frye FO], and  
Milekhin freeze-out on characteristic curve 
(\ref{v-epsilon}) 
[Milekhin FO]. 
}
\label{fig1e}
\end{figure}
\begin{figure}[thb]
\includegraphics[width=7.5cm]{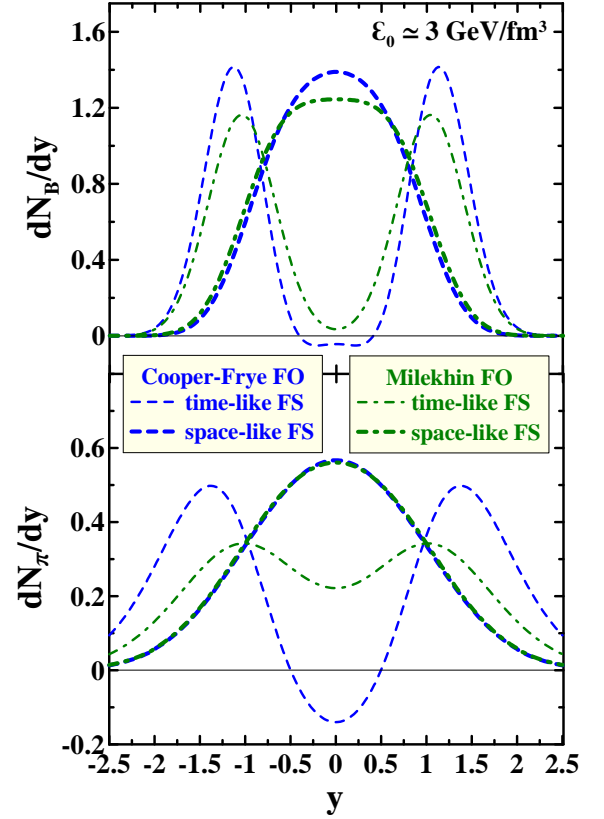}
\caption{(Color online)
Partial contributions to rapidity spectra of baryons (upper panel) and thermal pions  (lower panel) from space-like and time-like parts of the freeze-out surface. 
Spectra are calculated within two models of freeze-out: 
Cooper--Frye freeze-out on characteristic curve 
(\ref{v-epsilon})
[Cooper--Frye FO], and  
Milekhin freeze-out on the same characteristic curve 
[Milekhin FO]. 
}
\label{fig1f}
\end{figure}

It is of interest to compare particle spectra predicted by different models 
for freeze-out. In Fig. \ref{fig1e} rapidity distributions of various hadrons 
are demonstrated for three different prescriptions of the freeze-out: 
the above-described model of dynamical freeze-out (Dynamical FO)
[see Eqs. (\ref{FOcriterion1})--(\ref{FOspectrum})], 
Cooper--Frye freeze-out on characteristic curve (\ref{v-epsilon}) [Cooper--Frye FO], 
and Milekhin freeze-out on characteristic curve (\ref{v-epsilon}) [Milekhin FO]. 
These spectra were calculated within 1D hydrodynamic expansion of the 
slab of initial width of 4 fm and initial energy density 
$\varepsilon_0 = 3$ GeV/fm$^3$. Initial conditions 
for this slab are constructed on the assumption
that it is formed formed by the shock-wave mechanism
in head-on collision of two 1D slabs at
$E_{\scr{lab}}\simeq$ 10 $A$ GeV. Only nucleons, deltas and pions are included 
in the EoS used in this calculation. 
Displayed rapidity distributions are normalized to unit aria transverse 
to the direction ($x$) of longitudinal expansion. 
Freeze-out surfaces for these three models 
of freeze-out are displayed in Fig. \ref{fig0} for Dynamical FO and 
Fig. \ref{fig1d} (dashed curve for $\varepsilon = 0.4 \ \mbox{GeV/fm}^3$) 
for Cooper--Frye FO and Milekhin FO.

It is difficult to directly compare the Dynamical FO and Cooper--Frye FO, since 
the Dynamical FO surface is inappropriate for the Cooper--Frye FO and vise versa. 
This is because the Dynamical FO requires the hydrodynamics to be modified 
by removing the frozen-out matter, while the Cooper--Frye FO needs 
the hydro calculation to be run absolutely unrestricted. This is the reason why 
we consider the Milekhin FO, which takes place precisely on the same 
characteristic curve (\ref{v-epsilon}) as the Cooper--Frye FO. In particular, 
this is the reason why we compare only two models in Fig. \ref{fig1f}.

As seen from Fig. \ref{fig1e}, baryon rapidity distributions are quite 
close to each other within different freeze-out models. 
The Dynamical-FO and Cooper--Frye-FO are even strikingly close. 
Apparently, this 
occurs because they are confined by the baryon number conservation. 
This is not the case for thermal pions (i.e. without 
contribution of $\Delta$ decays). 
Difference between their distributions within different models is 
quite spectacular. 

As it is seen from Fig. \ref{fig1f}, this difference 
mainly originates from contributions from the time-like\footnote{We prefer to 
specify hypersurfaces by the character of space-time intervals within them, 
similarly to Refs. \cite{Bugaev96,Bugaev99} and contrary to Refs. 
\cite{Neumann97,Csernai97}}
parts of the freeze-out 
surface, i.e. from those parts, where the the Cooper--Frye 
normal vector $n^\mu_{CF}$ is space-like: $n_{CF}\cdot n_{CF} < 0$. 
Difference between Cooper--Frye and Milekhin recipes  is most strongly 
pronounced in this case. Note that the Cooper--Frye FO even reveals its generic 
deficiency---its "time-like" spectrum becomes negative at midrapidity. 
Physically it means that the midrapidity region is most abundantly populated 
by particles which have to be returned to the hydro phase. 
At the same time, contributions from the space-like parts
of the surface, where $n_{CF}\cdot n_{CF} > 0$, are quite similar 
within Cooper--Frye FO and Milekhin FO. Small diference between the 
Cooper--Frye normal time-like vector $n^\mu_{CF}$ and 
and the hydrodynamic 4-velocity results is tiny difference in 
"time-like" spectra.

Thus, spectra of newly produced particles are quite different in different 
models of freeze-out. At the same time, the numerical baryon-number 
and energy conservation is better than 1\% in all considered models. 
We still face the problem: which of these models is physically true. 
Unfortunately, above-mentioned experiments on 
expansion of compressed and heated classical fluid into vacuum
\cite{evap76,evap82,evap92,evap99}
cannot answer this question. As we have seen, particle number conservation
(baryon number, in Fig. \ref{fig1e}) makes these models fairly close 
to each other. We would like only to mention that the dynamical FO model 
naturally explains \cite{3FDpt} the  incident energy
behavior of inverse-slope parameters of transverse-mass spectra observed in
experiment. However, this is not a physical justification of this model. 
The question still remains open.

\begin{figure}[thb]
\includegraphics[width=7.9cm]{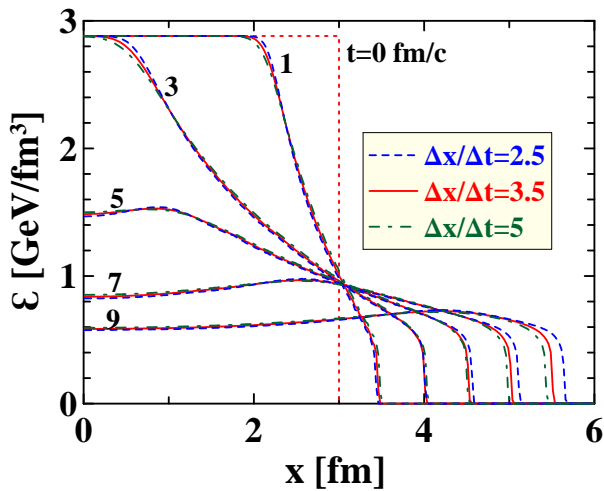}
\caption{(Color online)
The same as in Fig. \ref{fig1} (top panel)  but for different values of
$\Delta x /\Delta t$. Only results with freeze-out are displayed.
}
\label{fig1b}
\end{figure}

\begin{figure}[thb]
\includegraphics[width=7.9cm]{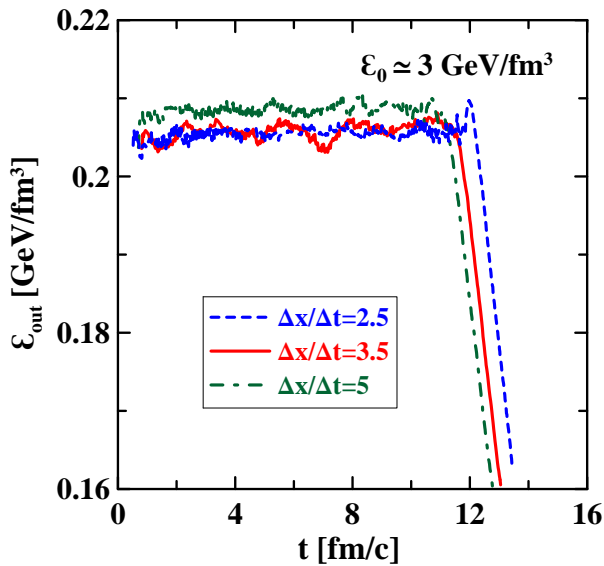}
\caption{(Color online)
The same as in Fig. \ref{fig1a}
(bottom panel) but for different values of
$\Delta x /\Delta t$.
}
\label{fig1c}
\end{figure}

Results of the above reported 1D simulations are stable with
respect to the numeric procedure \cite{Roshal81}.
They are quite insensitive to the size of the cell. 
The freeze-out front is smeared only over two cells independently of
their size and hence indeed is step-like in continuum limit.
Dependence on the most important numeric
parameter---the ratio of the space-grid step to the time step
$\Delta x /\Delta t$---is
displayed in Figs. \ref{fig1b} and \ref{fig1c}.
To reduce numerical diffusion, this ratio should be
taken optimal. As it was found in 1-dimensional simulations of
exactly solvable problems \cite{FCT-PIC}, the optimal range of
this ratio is $2.5<\Delta x /\Delta t<5$ with the preferable
$\Delta x /\Delta t \simeq 3.5$, minimizing the numerical
diffusion. As seen, within this range $2.5<\Delta x /\Delta t<5$
the results are really stable.

\subsection{3D simulations}
\label{3D simulations}

Condition ({\bf i}) (or Eq. (\ref{theta-frz})) ensures only that the
actual freeze-out energy density, at which the freeze-out actually
occurs, is less than $\varepsilon_{\scr{frz}}$.
\begin{figure}[thb]
\includegraphics[width=6.9cm]{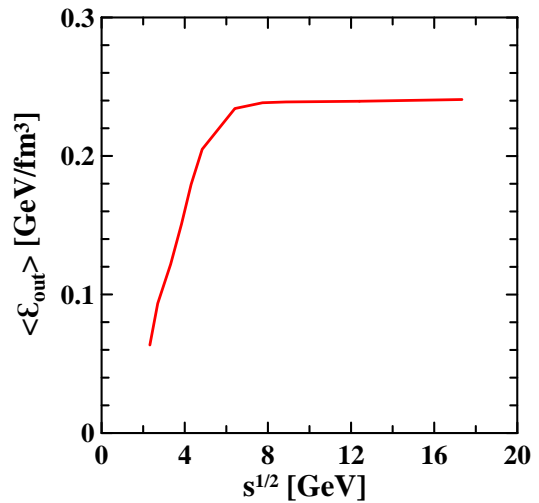}
\caption{(Color online)
Actual average freeze-out energy density
in central ($b=0$) Pb+Pb collisions as a function of
invariant incident energy.
}
\label{fig2a}
\end{figure}
\begin{figure*}[thb]
\includegraphics[width=17.cm]{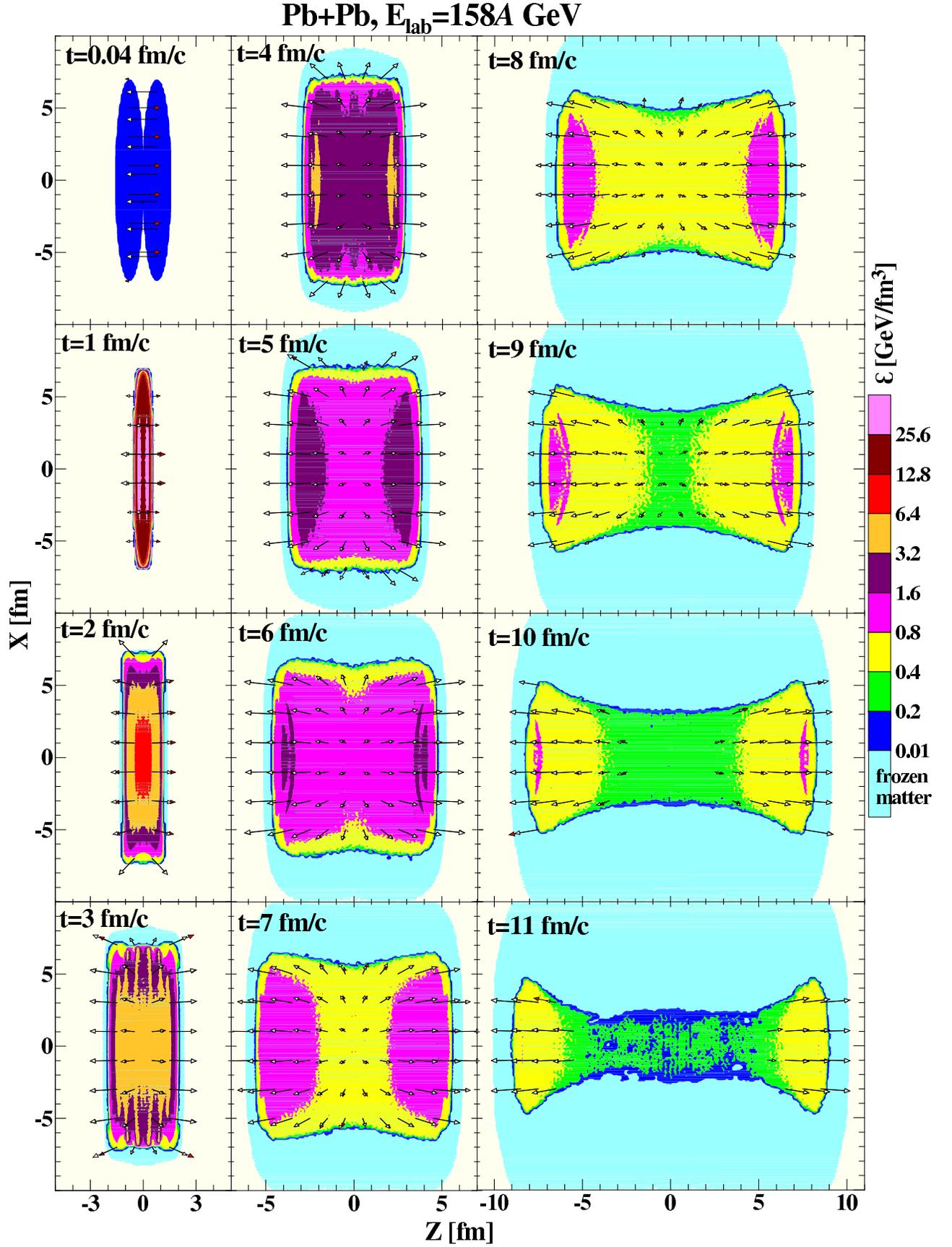}
\caption{(Color online) Time evolution (in the $xz$ plane of the
center-of-mass frame) of the total energy density (cf. Eq.
(\ref{eps_tot})) in the central Pb+Pb collision at $E_{\scr{lab}}=$
158$A$ GeV. The light-colored outer hallo corresponds to the
simulation without freeze-out and thus indicates the matter which
has already been frozen out to the time instant $t$. Arrows indicate
velocities of baryon-rich fluids: open arrows for the
projectile-like fluid and filled  arrows for the target-like fluid.
} \label{fig2c}
\end{figure*}
Therefore, $\varepsilon_{\scr{frz}}$ can be called a "trigger" value of
the freeze-out energy density.
As it has been demonstrated in the previous subsection,
a natural value
of this actual freeze-out energy density is $\varepsilon_{\scr{out}}
\approx \varepsilon^s/2$, i.e. at the middle of the fall from
$\varepsilon^s$ to zero. To find out the actual
value of $\varepsilon_{\scr{out}}$, we have to analyze results of
a particular simulation. In our previous paper \cite{3FD}
we have performed only a rough analysis of this kind.
This is why in the main text of Ref. \cite{3FD} we mentioned the value
of approximately 0.2 GeV/fm$^3$ for $\varepsilon_{\scr{out}}$ and in
the appendix
explained how the freeze-out actually proceeded\footnote{In terms of
  Ref. \cite{3FD} 
($\varepsilon_{\scr{frz[1]}}$ and 
$\varepsilon_{\scr{frz[1]}}^{\scr{code}}$)
our present quantities are 
$\varepsilon_{\scr{frz}}=\varepsilon_{\scr{frz[1]}}^{\scr{code}}$
and 
$\varepsilon_{\scr{out}}=\varepsilon_{\scr{frz[1]}}$.}.
Results of more comprehensive analysis for central ($b=0$) Pb+Pb
collisions are presented in Fig. \ref{fig2a},
which shows the  $\varepsilon_{\scr{out}}$
value averaged over space--time evolution of the collision:
$\langle\varepsilon_{\scr{out}}\rangle$. As seen,
$\langle\varepsilon_{\scr{out}}\rangle$ reveals saturation at the SPS
energies.
This happens in spite of the fact that our freeze-out condition
involves only a single constant parameter
$\varepsilon_{\scr{frz}}= 0.4$ GeV/fm$^3$, with the exception of low incident
energies, for which we use lower values: 
$\varepsilon_{\scr{frz}}(2A \mbox{ GeV}) = 0.3$ GeV/fm$^3$ and 
$\varepsilon_{\scr{frz}}(1A \mbox{ GeV}) = 0.2$ GeV/fm$^3$.

The "step-like" behavior of $\langle\varepsilon_{\scr{out}}\rangle$ is
a consequence of the freeze-out dynamics 
which has already been
illustrated in Fig. \ref{fig1a}. At low (AGS) incident energies,
the energy density achieved at the border with vacuum,
$\varepsilon^s$, is lower than
$\varepsilon_{\scr{frz}}$. The surface freeze-out stays at this
lower energy density up to
the global freeze-out because the freeze-out front cannot overcome the
supersonic 
barrier in the expanding matter, cf. the lower panel of Fig.
\ref{fig1}. 
At these low energies, the value
$\langle\varepsilon_{\scr{out}}\rangle$ turns out to be low
sensitive to the freeze-out parameter $\varepsilon_{\scr{frz}}$.
Only the global freeze-out {\bf (iiib)} of the system remnant, which
also contributes to  $\langle\varepsilon_{\scr{out}}\rangle$,
produces weak sensitivity to $\varepsilon_{\scr{frz}}$. The values 
$\varepsilon_{\scr{frz}}(2A \mbox{ GeV}) = 0.3$ GeV/fm$^3$ and 
$\varepsilon_{\scr{frz}}(1A \mbox{ GeV}) = 0.2$ GeV/fm$^3$ 
were precisely chosen in order to reduce contribution of the global
freeze-out to  $\langle\varepsilon_{\scr{out}}\rangle$.

With the incident energy rise the energy density achieved at the
border with vacuum gradually reaches the value of $\varepsilon_{\scr{frz}}$
and then even overshoot it. If the overshoot happens, the system
first expands without freeze-out. The freeze-out starts only when
$\varepsilon^s$ drops to the value of
$\varepsilon_{\scr{frz}}$. Then the surface freeze-out occurs really
at the value $\varepsilon^s \approx \varepsilon_{\scr{frz}}$
and thus the actual freeze-out energy density saturates at the value
$\langle\varepsilon_{\scr{out}}\rangle \approx \varepsilon_{\scr{frz}}/2$.

In Fig. \ref{fig2c} the time evolution of the total energy density
(cf. Eq. (\ref{eps_tot})) in the central Pb+Pb collision at
$E_{\scr{lab}}=$ 158$A$ GeV is displayed. The light-colored outer
hallo corresponds to the simulation without freeze-out and thus
indicates the matter which has already been frozen out to the time
instant $t$.  First of all we see that already in the beginning of
expansion stage ($t=2$ fm/c) the baryon-rich
fluids are mutually stopped and unified to a good extent, since
their hydrodynamic velocities almost coincide: arrows, originating
from the same point, are almost equal if not merged. This is not the
case for the baryon-free fluid, since its formation lasts till
approximately $t=4$ fm/c (see Fig. 18 in Ref. \cite{3FD}). At the
late stage of the expansion ($t\gsim 10$ fm/c) the baryon-rich and
baryon-free fluids become even spatially separated: The middle
region of the system, containing no arrows, is solely populated by  
the baryon-free fluid, and hence the baryon-rich matter falls into
two disconnected pieces. Thus, at the late stage of the evolution
the system effectively consists of three ``fireballs'' (two
baryon-rich and one baryon-free). This is in contrast to the
assumption of the statistical model
(\cite{Andronic06,Cleymans06,Randrup06,Dumitru06}),
where a single uniform ``fireball'' is considered. The
baryon-free ``fireball'' becomes frozen out first: the displayed time
instant $t=11$ fm/c is almost the last, when this ``fireball'' still
hydrodynamically evolves. 
Evolution of the two baryon-rich ``fireballs'' till the
complete freeze-out is rather long, till $t\approx 20$ fm/c. This
spatial separation of ``fireballs'' happens only at high incident
energies $E_{\scr{lab}}>$ 40$A$ GeV.

The freeze-out in the longitudinal direction proceeds accordingly to
the 1D pattern of Fig. \ref{fig1} (upper panel). 
In the transverse direction the freeze-out front moves
inwards the system. This a combined effect of fast longitudinal
expansion and comparatively slow transverse motion of the system.
This effect actually results in the ``two-fireballs'' structure at
the latest stage.

\begin{figure}[thb]
\includegraphics[width=8.4cm]{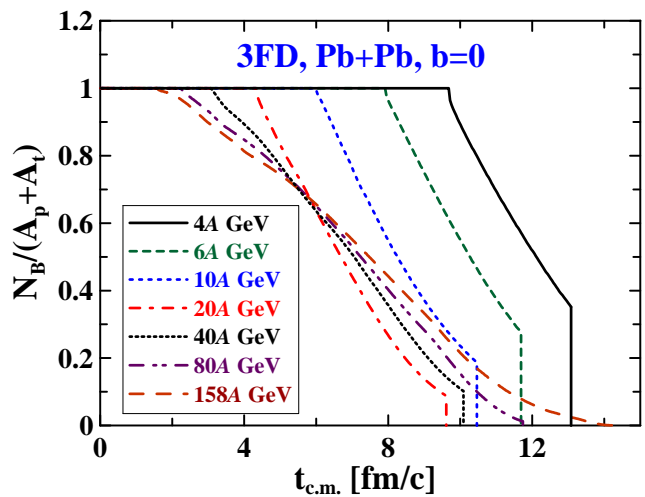}
\caption{(Color online)
Time evolution (in the center-of-mass frame)
of the non-frozen-out baryon
number (normalized to the total baryon number of the system
$A_p + A_t$) in central Pb+Pb collisions at various incident energies.
}
\label{fig2b}
\end{figure}

The occurrence of two spatially separated ``fireballs'' at the
latest stage of the collision is the reason why the global
freeze-out, cf. {\bf (iiib)}, does not occur in this system. At low
(AGS) incident energies, the surface freeze-out, actually occurring
at lower than $\varepsilon_{\scr{frz}}$ densities, takes place only
as long as the energy density in the center of expanding system
exceeds $\varepsilon_{\scr{frz}}$. When this center density drops
below $\varepsilon_{\scr{frz}}$, the rest of the system instantly
gets frozen out, as it is illustrated in Fig. \ref{fig2b} at the
example of time evolution the baryon number still involved in the
hydrodynamic phase. The abrupt fall of a curve corresponds to
the global freeze-out. In fact, only this global
freeze-out stage depends on the freeze-out parameter
$\varepsilon_{\scr{frz}}$ at low incident energies. With the
incident energy rise this instantly frozen-out remnant becomes
smaller and smaller until it completely disappears above 
$40A$ GeV energy.  This happens because outer freeze-out fronts stay
already at the $\varepsilon_{\scr{frz}}$ density, therefore the
maximum volume density can be only higher than
$\varepsilon_{\scr{frz}}$. At the same time the inner freeze-out
fronts overtake these two baryon-rich ``fireballs'' from behind and
result in gradual surface freeze-out of them without any remnant,
where criterion {\bf (iiib)} can work.

\section{Conclusions}

The described method of freeze-out can be called dynamical, since the
freeze-out process here is integrated into fluid dynamics through
hydrodynamic equations (\ref{eq1})--(\ref{theta-frz}). The freeze-out
front is not defined just ``geometrically'' on the condition of the
freeze-out criterion met but rather is a subject the fluid evolution.
It competes with the fluid flow and not always reaches the place where
the freeze-out criterion is met.

This kind of freeze-out is similar to the model of ``continuous
emission'' proposed in Ref. \cite{Sinyukov02}. There the particle emission
occurs from a surface layer of the mean-free-path width. In our case the
physical pattern is similar, only the mean free path is shrunk to zero.
In particular, the fact that the freeze-out sometimes occurs at lower
energy densities than that prescribed by
the freeze-out criterion can
be associated with the dependence on the future evolution of the fluid
system in the model of ``continuous
emission''. In that model the particle emission occurs from  interior of
the time-evolving system. Therefore, if the emitted particle is able to
leave the system or not depends on the expansion rate of
the system. Similar situations happen in our model. Sometimes
the freeze-out criterion is met too deep inside the system such that
rapid expansion of the system prevents the freeze-out front from
reaching this place.

We argue that the original Milekhin's method \cite{Milekhin} of
calculation of observable particle spectra is energy conserving, as
well as the Cooper--Frye recipe \cite{Cooper}, provided it is
considered on a discontinuous hypersurface,
i.e. that consisting of tiny (infinitely small in continuum
limit) fragments with normal vectors coinciding with the local
hydrodynamic 4-velocity.  
Moreover, we argue that Milekhin's approach on a discontinuous
hypersurface is natural for the Lagrangian formulation of hydrodynamics.
There are no
principal objections against using such a fragmented hypersurface
instead of the continuous one like in the Cooper--Frye method.
Discreetness of particle emission may hint at  discontinuous
character of the freeze-out hypersurface.

Systematic studies based on  hydrodynamic  models indicated that there
are two distinct  freeze-out points (see, e.g., \cite{Shuryak99,Heinz99}):
chemical and thermal ones.
Contrary to these studies
we do not need  two distinct  freeze-out points.
Multiplicities and
spectra are simultaneously and quite satisfactorily reproduced
with the single freeze-out
described above \cite{3FD}.
Of course,
we could somewhat refine reproduction of experimental data by introducing
two distinct freeze-out points. However, introduction of an additional
fitting parameter would be too high price for such a slight improvement.


\section*{Acknowledgements}


We are grateful to M.I. Gorenstein, E.E. Kolomeitsev, I.N. Mishustin, L.M. Satarov,
V.V.~Skokov, V.D. Toneev,  and D.N. Voskresensky for fruitful
discussions.
This work was supported by
Deutsche
Forschungsgemeinschaft (DFG project 436 RUS 113/558/0-3), 
Russian Foundation for Basic Research (RFBR grant 06-02-04001 NNIO\_a),
Russian Federal Agency for Science and Innovations
(grant NSh-8756.2006.2).

\appendix

\section{Problem of frozen-out hadrons returning into hydrodynamic phase}
\label{Problem}

When frozen-out matter coexists with still hydrodynamically evolving fluid, both 
the conventional Cooper--Frye recipe \cite{Cooper} and the model presented here 
suffer from a problem of frozen-out particles returning
to the hydro phase. For the Cooper--Frye method this problem was discussed in 
a number of papers \cite{Bugaev96,Neumann97,Csernai97,Bugaev99,Csernai04}. 
Here we would like to demonstrated that precisely the same 
(and  precisely to the same extent) problem exists for the freeze-out model 
discussed in this paper.

Let the fluid and the frozen-out matter be separated by a space surface with 
an external normal 3-vector  $\1 n$. Let this surface move with velocity 
$\1 v_\sigma$. Then we can decompose the number of particles frozen out 
on the element of hypersurface $d\sigma$, associated with the above separation 
surface, as follows (cf. Eq. (\ref{Cooper-FO}))
\begin{eqnarray}
\label{separation}
\hspace*{-5mm}
d N \!\!&=&\!\! d N_{\scr{esc.}} + d N_{\scr{ret.}} ,
\\
\label{escaping}
\hspace*{-5mm}
d N_{\scr{esc.}} \!\!&=& \!\!
 d\sigma \, n_\sigma^\mu \int\frac{d^3 p}{p_0} 
p_\mu  \, \ f (p,x) \,
\Theta ([\1 v - \1 v_\sigma] \cdot \1 n) ,
\\
\label{returning}
\hspace*{-5mm}
d N_{\scr{ret.}} \!\!&=&\!\!
 d\sigma \, n_\sigma^\mu \int\frac{d^3 p}{p_0} 
p_\mu  \ f (p,x) \,
\Theta ([\1 v_\sigma -\1 v ] \cdot \1 n) ,
\end{eqnarray}
where $\1 v = \1 p / p_0$ is the 3-velocity of a frozen-out particle. 
Here $d N_{\scr{esc.}} $ can be identified with a number of particles escaping 
from the system, since they move outward the system faster than the free surface 
overtakes them. While $d N_{\scr{ret.}}$ is a number of particles returning to 
the fluid. In the covariant form, $\Theta$-functions can be written as (see, e.g., 
Refs. \cite{Bugaev96,Neumann97,Csernai97,Bugaev99}) 
\begin{eqnarray}
\hspace*{-5mm}
\label{Theta-escaping}
\Theta ([\1 v - \1 v_\sigma] \cdot \1 n) &=&
\Theta (p_\mu  n_{CF\sigma}^\mu ),
\\
\label{Theta-returning}
\Theta ([\1 v_\sigma -\1 v ] \cdot \1 n) &=&
\Theta (-p_\mu  n_{CF\sigma}^\mu ),
\end{eqnarray}
where $n_{CF\sigma}^\mu$ is the normal 4-vector to the Cooper--Frye continuous 
hypersurface. This is always so, independently of the choice of $n_\sigma^\mu$ 
we use for calculation of spectra. Therefore, the fraction of returning particles, 
$d N_{\scr{ret.}} / d N_{\scr{esc.}}$ is independent of whether we employ 
the Cooper--Frye recipe with $n_\sigma^\mu=n_{CF\sigma}^\mu$ or the 
Milekhin's method with $n_\sigma^\mu=u^\mu$.


\begin{thebibliography}{999}
%
\bibitem{3FD}
 Yu.B. Ivanov, V.N. Russkikh, and V.D. Toneev,
 Phys. Rev. C {\bf 73}, 044904 (2006). 
%
\bibitem{Clare}
R.B. Clare and D. Strottman, Phys. Rept. {\bf 141} 177 (1986).
%
\bibitem{Stoecker86}
H. Stoecker and  W. Greiner, Phys. Rept. {\bf 137}, 277 (1986).
%
\bibitem{MRS91}
I.N.~Mishustin, V.N.~Russkikh, and
L.M.~Satarov, Yad. Fiz. {\bf 54}, 429 (1991) [Sov. J. Nucl. Phys.
{\bf 54}, 260 (1991)];
%
\bibitem{Rischke98}
D.H. Rischke, nucl-th/9809044.
%
\bibitem{Ruuskanen06}
P. Huovinen, P.V. Ruuskanen, nucl-th/0605008
%
\bibitem{3FDm}
V.D. Toneev, Yu.B. Ivanov, E.G. Nikonov, W. Norenberg, and V.N. Russkikh,
Phys. Part. Nucl. Lett. {\bf 2}, 288 (2005),
[Pi'ma o Fizike Elementarnykh Chastits i Atomnogo Yadra {\bf 2},
43 (2005)];
V.N. Russkikh, Yu.B. Ivanov, E.G. Nikonov, W. Norenberg, and
V.D. Toneev, Phys. Atom. Nucl. {\bf 67}, 199 (2004) [Yad. Fiz. {\bf  67}, 195 (2004)].
%
\bibitem{MRS88}  I.N.~Mishustin, V.N.~Russkikh, and
L.M.~Satarov, Yad. Fiz. {\bf 48}, 711 (1988) [Sov. J. Nucl. Phys.
{\bf 48},  454 (1988)];
%
Nucl. Phys. {\bf A494}, 595 (1989).
%
\bibitem{INNTS}
      Yu.B.~Ivanov, E.G.~Nikonov, W.~N\"orenberg, V.D.~Toneev, and
      A.A.~Shanenko,  Heavy Ion Phys. {\bf 15}, 127 (2002).
%
\bibitem{Kat93}
U.~Katscher, D.H.~Rischke, J.A.~Maruhn, W.~Greiner,
I.N.~Mishustin, and L.M.~Satarov, Z. Phys. {\bf A346}, 209 (1993);
%
U.~Katscher, J.A.~Maruhn, W.~Greiner, and  I.N.~Mishustin, Z.
Phys. {\bf A346}, 251 (1993);
%
A.~Dumitru, U.~Katscher, J.A.~Maruhn, H.~St\"ocker, W.~Greiner,
and D.H.~Rischke, Phys. Rev.  C {\bf 51}, 2166 (1995);
%
Z. Phys. {\bf A353}, 187 (1995). 
%
\bibitem{Brac97}
J.~Brachmann, A.~Dumitru, J.A.~Maruhn, H.~St\"ocker,
W.~Greiner, and D.H.~Rischke, Nucl. Phys. {\bf A619}, 391 (1997);
%
A.~Dumitru, J.~Brachmann, M.~Bleicher, J.A.~Maruhn, H.~St\"ocker,
and W.~Greiner, Heavy Ion Phys. {\bf 5}, 357 (1997);
%
M.~Reiter, A.~Dumitru, J.~Brachmann, J.A.~Maruhn, H.~St\"ocker,
and W.~Greiner, Nucl. Phys. {\bf A643}, 99 (1998);
%
M.~Bleicher, M.~Reiter, A.~Dumitru, J.~Brachmann, C.~Spieles,
S.A.~Bass, H.~St\"ocker,  and W.~Greiner, Phys. Rev. C {\bf 59},
R1844 (1999); 
%
J.~Brachmann, A.~Dumitru, H.~St\"ocker, and W.~Greiner,
  Eur. Phys. J. {\bf A8}, 549 (2000); 
%
\bibitem{Brac00a}
J.~Brachmann, S.~Soff, A.~Dumitru, H.~St\"ocker, J.A.~Maruhn,
W.~Greiner, L.V.~Bravina, and D.H.~Rischke, Phys. Rev. C {\bf 61},
024909 (2000). 
%
\bibitem{3FDflow}
V.N. Russkikh and Yu.B. Ivanov,
Phys. Rev. C {\bf 74}, 034904 (2006).
%
\bibitem{3FDpt}
Yu.B. Ivanov and V.N. Russkikh, nucl-th/0607070
%
\bibitem{gasEOS}
V.M. Galitsky and I.N. Mishustin, Sov. J. Nucl. Phys. {\bf 29}, 181
(1979).
%
\bibitem{E866}
L. Ahle, {\it et al.},
 Phys. Lett. {\bf B476}, 1 (2000).
%
\bibitem{NA49}
 S. V. Afanasiev,  {\it et al.},
 Phys. Rev. C {\bf 66},054902 (2002);
C. Alt,  {\it et al.},
J. Phys. {\bf G30}, S119 (2004);
M. Gazdzicki {\it et al.},
Phys. {\bf G30}, S701 (2004).
%
\bibitem{Gorenstein03}
M.I. Gorenstein, M. Gazdzicki, and K. Bugaev,
Phys. Lett. {\bf B567},  175 (2003).
%
\bibitem{Mohanty03}
B. Mohanty, J. Alam, S. Sarkar,
T.K. Nayak, B.K. Nandi,  Phys. Rev. C {\bf 68}, 021901 (2003).
%
\bibitem{Bratkovskaya}
E.L. Bratkovskaya, M. Bleicher, M. Reiter, S. Soff, H. Stoecker,
M. van Leeuwen, S. Bass, and  W. Cassing,
Phys. Rev. C {\bf 69}, 054907 (2004);
E.L. Bratkovskaya, S. Soff, H. Stoecker, M. van Leeuwen, and W. Cassing,
Phys. Rev. Lett. {\bf 92}, 032302 (2004).
%
\bibitem{Hama04}
M. Gazdzicki, M.I. Gorenstein, F. Grassi, Y. Hama, T. Kodama, and O. Socolowski Jr,
 Braz. J. Phys. {\bf 34}, 322 (2004).
%
\bibitem{Milekhin} G.A. Milekhin, Zh. Eksp. Teor. Fiz. {\bf 35}, 1185 (1958);
  Sov. Phys. JETP {\bf 35}, 829 (1959); Trudy FIAN {\bf 16}, 51 (1961).
%
\bibitem{Cooper}  F. Cooper and G. Frye, Phys. Rev. D {\bf 10}, 186 (1974).
%
\bibitem{Ris95a}
        D.H. Rischke, Y. P\"urs\"un, J.A. Maruhn, H. St\"ocker,
        W. Greiner, Heavy Ion Phys. {\bf 1}, 309 (1995).
%
\bibitem{Kol99}
J.~Sollfrank, P.~Huovinen, M.~Kataja,
  P.V.~Ruuskanen, M.~Prakash, and R.~Venugopalan, Phys. Rev. C {\bf  55},
  392 (1997); P.~Huovinen,  P.V.~Ruuskanen and  J.~Sollfrank,
  Nucl. Phys. {\bf A650}, 227 (1999); P.F. Kolb, J. Sollfrank,
  P.V. Ruuskanen, and U. Heinz, Nucl. Phys. {\bf A661}, 349 (1999);
         P.F. Kolb, J. Sollfrank, U. Heinz, Phys. Lett.
         {\bf B 459}, 667  (1999);
P.F. Kolb, P. Huovinen, U. Heinz,
         H. Heiselberg, Phys. Lett. {\bf B 500}, 232 (2001).
%
\bibitem{HS95}
C.M.~Hung and  E.V.~Shuryak,  Phys. Rev. Lett. {\bf 75},
4003 (1995); C.M.~Hung and  E.~Shuryak,   Phys. Rev. C {\bf 57}, 1891
(1998).
%
%
\bibitem{Teaney01} 
D.~Teaney, J.~Lauret, and E.V.~Shuryak, nucl-th/0110037; 
Phys. Rev. Lett. {\bf 86}, 4783 (2001).
%
\bibitem{Hirano07} 
T. Hirano, U.W. Heinz, D. Kharzeev, R. Lacey, and Y. Nara, 
 arXiv:nucl-th/0701075
%
\bibitem{Bass00}
A. Dumitru, S.A. Bass, M. Bleicher, H. Stocker, and W. Greiner,
Phys. Lett. {\bf B460}, 411, (1999);
S.A. Bass, A. Dumitru, M. Bleicher, L. Bravina, E. Zabrodin, H. Stoecker,
and W. Greiner,
Phys. Rev. C {\bf 60}, 021902, (1999);
S.A. Bass and A. Dumitru, Phys. Rev. C {\bf 61}, 064909, (2000).
%
\bibitem{Per00}
         D.Yu. Peressounko and Yu.E. Pokrovsky,
         Nucl. Phys. {\bf A669}, 196 (2000).
%
\bibitem{Non00}
        C. Nonaka, E. Honda, S. Muroya,
        Eur. Phys. J. {\bf C17}, 663 (2000).
%
\bibitem{Hir02}
        T. Hirano, Phys. Rev. C 65 (2002) 011901(R);
        T. Hirano and K. Tsuda, Phys. Rev. C {\bf 66}, 054905 (2002).
%
\bibitem{Ham05}
 Y. Hama, T. Kodama, and O. Socolowski, Braz. J. Phys. {\bf 35}, 24
 (2005).
%
\bibitem{Nonaka06}
C. Nonaka and S.A. Bass, Nucl. Phys. {\bf A774}, 873 (2006);
nucl-th/0607018.
%
\bibitem{Hirano06}
T. Hirano, U.W. Heinz, D. Kharzeev, R. Lacey, and Y. Nara,
Phys. Lett.  {\bf B636}, 299, (2006).
%
\bibitem{Sat06}
L.M. Satarov, A.V. Merdeev, I.N. Mishustin, and H. Stocker,
hep-ph/0606074, hep-ph/0611099
%
\bibitem{Bugaev96}  K.A. Bugaev, Nucl. Phys. {\bf A606}, 559 (1996).
%
\bibitem{Neumann97}  J.J. Neumann, B. Lavrenchuk, and G. Fai, Heavy Ion
  Physics {\bf 5}, 27 (1997).
%
\bibitem{Csernai97}  L.P. Csernai, Z. L\'az\'ar, and D. Moln\'ar,
Heavy Ion Phys. {\bf 5}, 467 (1997).
%
\bibitem{Bugaev99}  K.A. Bugaev and M.I. Gorenstein, nucl-th/9903072;
%
K.A. Bugaev,  M.I. Gorenstein, and W. Greiner, J.Phys.{\bf G25}, 2147
(1999); 
%
Heavy Ion Phys. {\bf 10}, 333 (1999). 
%
\bibitem{Csernai99} Cs. Anderlik, L.P. Csernai, F. Grassi, Y. Hama,
  T. Kodama, Zs. L\'az\'ar, and H. St\"ocker, Heavy Ion Phys. {\bf 9},
  193 (1999). 
%
\bibitem{Csernai04} K. Tamo\v{s}i\={u}nas and   L.P. Csernai,
  Eur. Phys. J. {\bf A20}, 269 (2004). 
%
\bibitem{Weinberg}
Steven Weinberg, {\em ``Gravitation and Cosmology: Principles and
  Applications of the General Theory of Relativity''} (John Wiley and
Sons, New York, 1972), Chapter 2, sects. 6 and 8.
%
\bibitem{Gorenstein84} M.I. Gorenstein and Yu.M. Sinyukov,
Phys. Lett. {\bf B142}, 425 (1984).
%
\bibitem{Csernai05}
E. Molnar, L. P. Csernai, V. K. Magas, A. Nyiri, and
K. Tamo\v{s}i\={u}nas, Phys. Rev C {\bf 74}, 024907 (2006);
%
\bibitem{Sinyukov02}
F. Grassi, Y. Hama, and T. Kodama, Phys. Lett. {\bf B355}, 9 (1995);
Z. Phys. {\bf C73}, 153 (1996);
Yu.M. Sinyukov, S.V. Akkelin, and Y. Hama, Phys. Rev. Lett. {\bf 89},
052301 (2002);
%
F. Grassi, Braz. J. Phys. {\bf 35}, 52 (2005).
%
%
%
%
\bibitem{Land-Lif}
L.D. Landau and E.M. Lifshitz, {\em ``Fluid Mechanics''} (Pergamon
Press, Oxford, 1979).
%
%
%
%
\bibitem{evap76}
C.J. Knight,  
J. Fluid Mech. 
{\bf 75}, 469 (1976)
%
%
%
%
\bibitem{evap82}
J.E. Shepherd and B. Sturtevant, 
J. Fluid Mech. 
{\bf 121}, 379 (1982).
%


%
%
%
\bibitem{evap92}
Th. Kurschat, H. Chaves and G.E.A. Meier 
J. Fluid Mech. 
{\bf 236}, 43 (1992).
%
%
%
%
\bibitem{evap99}
J.R. Simoes-Moreira and J.E. Shepherd,  
J. Fluid Mech. {\bf 382}, 63 (1999). 
%
%
%
%
%
%
%
\bibitem{Roshal81}
A.S. Roshal and V.N. Russkikh, Yad. Fiz. {\bf 33}, 1520 (1981).
%
\bibitem{FCT-PIC}   
V.N. Russkikh, in ``Numerical Methods of Medium Mechanics'' (in
Russian), Novosibirsk, vol. {\bf 1(18)} 104 (1987).
%
\bibitem{Harlow76}
    F.H.  Harlow, A.A. Amsden, and J.R. Nix,
J. Comp. Phys. {\bf 20}, 119 (1976).
%
\bibitem{Andronic06}
A. Andronic, P. Braun-Munzinger, and J. Stachel, Nucl. Phys. {\bf A772}, 167 (2006);
%
\bibitem{Cleymans06}
J. Cleymans, H. Oeschler, K. Redlich, and S. Wheaton,
 Phys. Rev. C {\bf 73}, 034905 (2006); 
hep-ph/0607164
%
\bibitem{Randrup06}
J. Randrup and J. Cleymans,
 Phys. Rev. {\bf C 74}, 047901 (2006)
%
\bibitem{Dumitru06}
A. Dumitru, L. Portugal, and D. Zschiesche,
Phys. Rev. C {\bf 73},  024902 (2006). 
%
\bibitem{Shuryak99}
E.V. Shuryak, Nucl. Phys. {\bf A661}, 119c (1999).
%
\bibitem{Heinz99}
U. Heinz, Nucl. Phys. {\bf A661}, 141c (1999).



%
\end{thebibliography}
\end{document}